\documentclass[manuscript, nonacm]{acmart}

\usepackage{tikz}
\usepackage{amsmath}
\usepackage{amsfonts}
\usepackage{graphicx}
\usepackage{subcaption}
\usepackage{stackengine}
\usepackage{xcolor}
\usepackage{array, booktabs, makecell}
\usepackage{bm}
\usepackage{balance}
\usepackage[british]{babel} 

\def\delequal{\mathrel{\ensurestackMath{\stackon[1pt]{=}{\scriptstyle\Delta}}}}

\usepackage{abstract} 

\definecolor{lblu}{rgb}{0.74, 0.83, 0.9}
\definecolor{dgre}{rgb}{0.56, 0.74, 0.56}
\definecolor{iora}{rgb}{1.0, 0.31, 0.0}

\newcommand{\ie}{\mbox{{i.e.,\ }}}

\newcommand{\para}[1]{\vspace*{1ex}\noindent\textbf{#1} }
\newcommand{\bx}{\bm{x}}%
\newcommand{\bz}{\bm{z}}%
\newcommand{\etheta}{e_{\theta}}
\newcommand{\dgamma}{d_{\gamma}}
\newcommand{\Norm}[1]{\left\lVert#1\right\rVert}

\begin{document}

\title{Zooming Into the Darknet: 
Characterizing Internet Background Radiation and its Structural Changes}


\author{Michalis Kallitsis}
 \affiliation{%
   \institution{Merit Network, Inc.}
   \country{USA}}
\email{mgkallit@merit.edu}

\author{Vasant Honavar}
 \affiliation{%
   \institution{Penn State University}
   \country{USA}}
\email{vuh14@psu.edu}

\author{Rupesh Prajapati}
 \affiliation{%
   \institution{Penn State University}
   \country{USA}}
\email{rxp338@psu.edu}

\author{Dinghao Wu}
 \affiliation{%
   \institution{Penn State University}
   \country{USA}}
\email{dux12@psu.edu}

\author{John Yen}
 \affiliation{%
   \institution{Penn State University}
   \country{USA}}
\email{juy1@psu.edu}

\maketitle

\begin{abstract}
Network telescopes or ``Darknets" provide a unique window into Internet-wide malicious 
activities associated with 
malware propagation, denial of service attacks,  
scanning performed for network reconnaissance, and others. 
Analyses of the resulting data can 
provide actionable insights to security analysts that can be used to 
prevent or mitigate cyber-threats. 
Large Darknets, however, observe millions of nefarious events
on a daily basis which makes the transformation of the captured 
information into meaningful insights challenging. 
We present a novel framework for characterizing Darknet behavior
and its temporal evolution aiming to address this challenge. 
The proposed framework: (i) Extracts a high dimensional representation of Darknet
events composed of features distilled from Darknet data and other external sources; 
(ii) Learns, in an unsupervised fashion, 
an information-preserving low-dimensional representation of these events 
(using deep representation learning) that is amenable to clustering; 
(iv) Performs clustering of the scanner data in the resulting representation space
and provides interpretable insights using optimal decision trees; and
(v) Utilizes the clustering outcomes as ``signatures" that can be used
to detect structural changes in the Darknet activities. We evaluate the proposed
system on a large operational Network Telescope and demonstrate its ability
to detect real-world, high-impact cybersecurity incidents. 
\end{abstract}

\section{Introduction}

Cyber-attacks present one of the most severe threats to 
the safety of citizenry and the
security of the nation's
critical infrastructure (e.g., energy grid, transportation network, health system, 
food and water supply networks).
A critical phase in cyber-attack is ``reconnaissance", which often
involves ``scanning" for potentially vulnerable machines or devices on the internet 
so that that these vulnerabilities may be exploited in later phases of the 
cyber-attack.
Similarly, malware that attempt
to propagate from one compromised machine to other
vulnerable devices are also engaged in malicious
scanning activities. Such actions are difficult to be identified
in an operational network because they are oftentimes
low-volume and interwoven with normal network traffic,
behaving similarly lest they are detected.

Characterizing these scanning behaviors can provide important information
for network security analysts because they may reveal change 
of attack strategies, new vulnerabilities that are being exploited,
and unauthorized use of Internet resources.
Network telescopes~\cite{caida_telescope_report}, also known as ``Darknets", provide
a unique opportunity for characterizing and detecting Internet-wide
malicious activities.
A Darknet receives and records unsolicited traffic---coined as 
Internet Background Radiation (IBR)---destined to an \emph{unused} but \emph{routed} address space.
This ``dark IP space" hosts no services or devices, and therefore any traffic arriving to it is inherently malicious. No regular user traffic reaches the Darknet. Thus, Darknets
have been frequently used by the networking and security communities
to shed light into dubious malware propagation and interminable network scanning activities~\cite{
Durumeric:2014:IVI:2671225.2671230,
Wustrow2010IBRrevisted,Pang2004IBR_active_responses,Iglesias2017IBRpatterns,ScannersDarknet2019}.
They have also been used to detect cyber-threats, e.g., botnets \cite{MiraiUSENIX2017}, DDoS and other types of attacks \cite{Fachkha2015DDoS, Ban2012LongTermAttack,Nishikaze2015TAP}, and to detect novel attack patterns \cite{Ban2016MiningClustering}.

Our team has access to a large network telescope for gathering observations of  
scanning activity across the Internet. The resulting data can be used to characterize these 
scanning events, and to gain actionable insights for preventing or mitigating such cyber-threats. 
An important task in this context has to do with \emph{characterizing the different 
Darknet scanners},
based on their traffic profile, the characteristics of their targets, 
their port scanning patterns, etc., and then employing these
characterizations as signatures to \emph{detect temporal changes
in the evolution of the Darknet}.
This problem can be reformulated as a problem of 
unsupervised clustering. However, the resulting clustering problem presents several
non-trivial challenges: (i) The data are heterogeneous with regard to the types of observations included. 
For example, some of the observations are categorical while others are numeric. However, 
standard clustering methods are typically designed to handle a single type of data, which 
renders them not directly applicable to the problem of clustering the scanner data; 
(ii) The number of observed variables, e.g., the  ports scanned over the duration of monitoring,  for each scanner can be in the order of thousands, 
resulting in extremely high-dimensional data.  Distance calculations are known to be inherently
unreliable in high-dimensional settings \cite{aggarwal2001surprising}, making it challenging to apply 
standard clustering methods that rely on measuring distance between data samples to cluster 
them; (iii) Linear dimensionality reduction techniques  such
as Principal Component Analysis (PCA) \cite{jolliffe2016principal} fail to cope with non-linear 
interactions between the observed variables;
(iv) interpreting the clustering outcome, that may include
hundreds of clusters with high-dimensional features, can be a non-trivial task.  

Against this background, this paper explores a novel unsupervised approach to characterizing 
network scanners using observations from a Network Telescope, that overcomes the aforementioned 
challenges. Our system (see Figure~\ref{fig:orion})
starts with \emph{unstructured, raw} packet data collected in our Darknet,
identifies \emph{all} Darknet scanning events within the monitoring interval of our interest,
\emph{annotates} these events with external data sources such as routing, DNS, geo-location
and data from \texttt{Censys.io}~\cite{censys}, distills an array of features to \emph{profile the behavior of each scanner},
and passes the set of feature-rich scanners to an unsupervised clustering method. The output of clustering is a grouping of the scanners into a number of clusters based on their scanning profiles. 

The key contributions of the paper are as follows:
We leverage the recent advances in deep neural networks (DNN) 
and employ powerful ``embedding" or ``representation learning" methods \cite{bengio2013representation} 
to automate the construction of low-dimensional vector space representations of heterogeneous, 
complex, high-dimensional network scanner data.  We apply standard 
clustering methods e.g., $K$-means, to the resulting information-preserving embedding of the data. 
We then introduce the use of \emph{optimal classification trees} as a means
for interpreting the clustering outcome and for understanding the structure of the
underlying data. To automate the detection of structural changes in the
Darknet, we propose the use of a \emph{Wasserstein metric} to help us assess
the similarity or ``distance`` between the clustering outcomes of interest (e.g.,
comparing Darknet clusters between consecutive days). 
Finally, we evaluate the proposed framework on real-world Darknet activities that span a period of three months.

The rest of the paper is organized as follows: 
Section \ref{sec:related} describes existing work related 
to this research. 
Section \ref{sec:DarknetObservabilities} describes the
Network Telescope data used in our work and the pipeline
for enriching them with additional data sources,
e.g., geolocation, data from \texttt{Censys.io}, etc.
Section~\ref{sec:clusteringScanners} describes our approach to learning low-dimensional, 
information-preserving representations of the resulting high-dimensional data,
which is then used for clustering Darknet events.
In Section \ref{sec:peava} we discuss tuning and evaluation
of the proposed deep learning tasks.
Section~\ref{sec:odt} introduces the usage of decision trees 
for clustering interpretation and in Section~\ref{sec:emd}
we discuss the ``Earth Mover's Distance" formulation
used as the Wasserstein metric for assessing inter-cluster similarity
and helping with ``novelty detection" of temporal patterns.
Section \ref{sec:real} presents case-studies and insights 
from real-world events detected with the proposed approach
and we conclude 
the paper with a brief summary of our contributions 
along with some directions for future research.

\section{Related Work}
\label{sec:related}


As noted above,  Darknets provide a unique perspective
into Internet-wide scanning activities involved in malware propagation,  network reconnaissance, Denial of Service (DoS) attacks, misconfigurations, etc.
\cite{
Durumeric:2014:IVI:2671225.2671230,jonker2017millions,Dainotti2014botnetIBR,Wustrow2010IBRrevisted,Herwig2019HajimeIRB,ScannersDarknet2019}. 
Darknet data have been utilized
to study DDoS attacks \cite{Czyz2014NTPDDoS,Fachkha2015DDoS,chiang2000fault,Wang2018DDoS},
DoS attacks and BGP blackholing \cite{Jonker2018DoS_BGP_IBR},
IPv6 routing instabilities \cite{Czyz2013IPv6IBR},
and long-term cyber attacks \cite{Ban2012LongTermAttack}. 
Application-level responses to IBR observed in Darknet
have also been used to characterize Internet-wide 
scanning activities
\cite{Pang2004IBR_active_responses}.


Of particular interest in this context is the use of Darknet data for 
detecting and characterizing new malware.
The Mirai botnet, for instance, is known to have started its malware propagation activity
by first scanning port 23 (Telnet) for potential victims in
the Internet \cite{MiraiUSENIX2017}.  Over time, 
and as  Internet-of-Things (IoT) devices had proven
to be very susceptible in getting 
compromised by malware infection, its scanning behavior changed as well.  
It proceeded to  scan
port 2323, and eventually
10 other ports \cite{MiraiUSENIX2017}. 
Hence, reliably detecting and responding to such attacks calls for effective methods 
for rapid identification of novel signatures of  malware behavior. 
Volumetric attacks on a target, e.g., as part of a distributed DoS attack through IP 
spoofing, i.e., forging the source IP address of a packet, 
are also identifiable in a Darknet and referred as \emph{backscatter}~\cite{moore01inferring}.
Recent work has shown that thousands of victims of DoS attacks can be identified by analyzing Darknet \emph{backscatter} data~\cite{jonker2017millions,Iglesias2017IBRpatterns}. 

Clustering offers a powerful approach to the analyses of Darknet data to identify novel attack patterns, victims of attacks, novel network scanners, etc. \cite{Nishikaze2015TAP,Ban2016MiningClustering,Iglesias2017IBRpatterns}.
For example, Nishikaze et al. ~\cite{Nishikaze2015TAP} encode Darknet traffic using 
27 network features
associated with blocks / subnets
of the IP space and cluster the resulting data to cluster the subnets according to their traffic profiles. Ban et al. \cite{Ban2016MiningClustering} have shown how clustering of Darknet data, followed by frequent pattern mining and visualization can be used to detect novel attack patterns. Iglesias and Szeby \cite{Iglesias2017IBRpatterns} have shown how to cluster IBR data from Darknet based on a novel representation of network traffic to identify network traffic patterns that are characteristic of activities such as long term scanning, as well as bursty events from targeted attacks and short term incidents.
Finally, Sarabi and Liu~\cite{10.1145/3278532.3278545} employ
deep learning for obtaining lightweight embeddings to characterize the population of Internet hosts
as observed by scanning services such as \texttt{Censys.io}.

\section{Network Telescope Data}
\label{sec:DarknetObservabilities}

Network telescopes offer a unique vantage point into macroscopic Internet-wide activities.
Specifically, they offer the ability 
to detect a broad range of dubious scanning activities;
from high-intensity scanning to 
low-speed, seemingly innocuous nefarious behaviors, which are much harder to detect in a large-scale operational network.  
Typical approaches to detecting scanning 
in an operational network set a (somewhat arbitrary) threshold on the number of packets received from a suspicious host within a time period
or a threshold on the number of unique destinations contacted by the host 
(e.g., 25 unique destinations with 5 minutes) 
as the detection criterion for suspected malicious behaviors.  While this
approach can indeed catch some dubious activities, it fails to capture those that occur at a frequency that is below the set threshold.  On the other hand, lowering the threshold would inevitably include many more non-malicious events, hence overwhelming the analysts (i.e., high-alert ``fatigue")
and significantly increase the complexity of
further analyses aiming at distinguishing malicious 
events from normal ones.  
Because benign real-user
network traffic does not reach the Darknet, scanning activities gathered at
the Darknet do not need to be filtered, thus obviating the need to set an arbitrary threshold.  Hence, even low-speed malicious activities can be easily detected in a Darknet that is sufficiently large~\cite{caida_telescope_report}.

\begin{figure}[t]
    \centering
    \includegraphics[width=1\linewidth,trim=3em 0 2.9em 3em,clip]{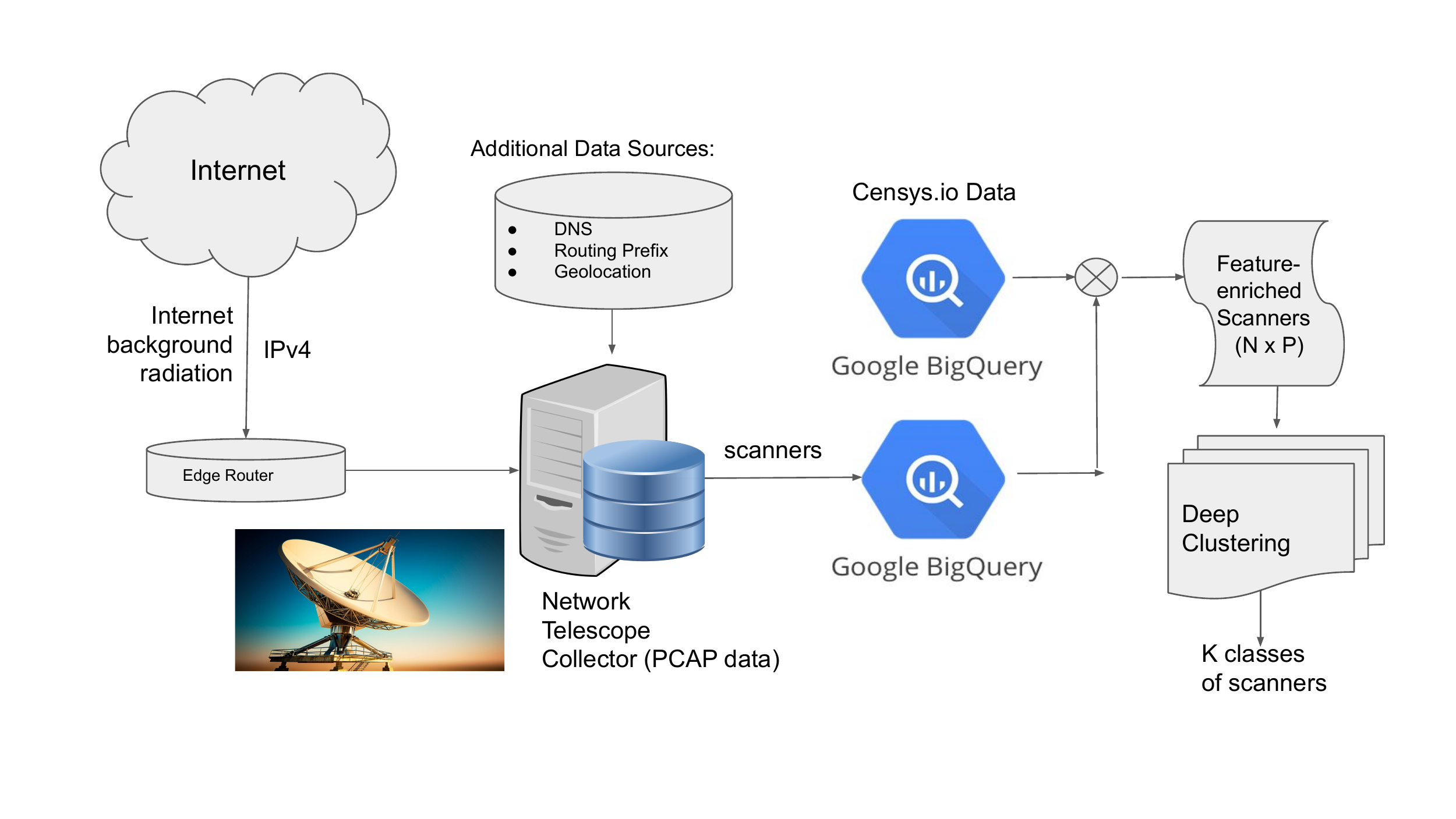}
    \vspace{-40pt}
    \caption{The Darknet data infrastructure.}
    \label{fig:orion}
\end{figure}

In this section we describe the networking and processing instrumentation
that provide us with a near-real-time pipeline for extracting and annotating scanners.
Packets arriving in our /13 dark IP space are collected in \texttt{PCAP} format on an hourly basis.
During a typical day, more than 100 GB of \emph{compressed} Darknet data is collected consisting of some
3 billion packets on average. As Figure~\ref{fig:orion} depicts, 
the PCAP data is processed to extract Darknet events (such
as scanning and backscatter)
and annotate them with external
data sources such as DNS, geolocation information
using the MaxMind databases~\cite{GeoIP} and routing information from CAIDA's prefix to AS mapping
dataset~\cite{pf2as}. A Darknet event is identified by 
i) the observed source IP, the ii) protocol flags used and iii) the targeted port.
We employ caching to keep ongoing scanners and other events in memory.
When an event remains 
inactive for a period of about 10 minutes, it ``expires" from the cache and gets recorded
to disk. Note here that scanners that target multiple ports and/or protocols would be
tracked in multiple separate events. This is a key observation and plays an 
important role in the construction of scanning features that we will introduce in the sequel.

All identified Darknet events are also uploaded in near-real-time to Google's BigQuery~\cite{bq}
for efficient analysis, further processing and also ease of data sharing. In addition, storing the extracted
events into BigQuery tables enables easy integration with extra data sources
also available in BigQuery, namely \texttt{Censys.io} data~\cite{censys, Durumeric:2015:SEB:2810103.2813703}.
Censys actively scans the whole IPv4 space and their 
data provide a unique perspective on the nature of a scanner since they potentially include information
about the open ports and services at the scanning host itself. Such coupling of information
helped the authors in~\cite{MiraiUSENIX2017} to identify device types and manufacturer information
of the infected Mirai population. We use Censys data in a similar manner and employ them
to enrich the scanner features we use for our clustering task.
The Darknet data and integrated data sources are aggregated on a \emph{daily basis}
to construct clustering features for the Darknet events, as described in the next section.

\section{Clustering Darknet Events}
\label{sec:clusteringScanners}

We designed the feature set having the following two
considerations in mind:
First, the dimensionality of the feature space  
is very high (i.e., in the order of thousands). 
Second, the evaluation and interpretation
of the clustering results of scanners could be challenging because 
we have no ``ground truth" or clustering labels.
One therefore needs to use semantics extracted from the
data itself.  In this section,
we describe our approach for addressing the high dimensionality challenges through a combination of
1) one-hot encoding of high-dimensional features (e.g., ports), 
and 2) deep learning for extracting a
low-dimension latent representation. Section~\ref{sec:peava} will describe our methodology for
clustering evaluation.

\subsection{Feature Design for Darknet Events}
\label{sec:features}

Table~\ref{tab:all_features} summarizes the features
used to represent Darknet activity. 
As mentioned earlier, Darknet events in this study are aggregated on a daily basis.
For example, a scanner identified $n$ times within a given day---possibly 
scanning multiple ports and protocols---will be represented
in the clustering input as a single entry annotated with the features of Table~\ref{tab:all_features}.
The chosen features fall into the following categories:
1)  intensity of the activity (e.g., total number of packets captured
in the Darknet for a given scanner within the day, total duration of scanning, number
of ports scanned, etc.), 
2) scanning intentions (i.e., the set of ports and protocols involved),
3) scanning strategy (i.e., information about the unique destinations
reached in the Darknet, both in terms of IPs and /24 prefixes),
and 4)
external threat intelligence for the scanner itself (i.e., information gathered from \texttt{Censys.io}).
We also annotate each event with BGP routing information, organizational and geolocation features,
DNS-based features and others to help the analyst interpret the clustering results.

Notice here that several features used are extremely high
dimensional; e.g.,
the number of unique TCP/UDP ports is $2^{16}$. 
Therefore, we chose an \emph{one-hot encoding} scheme for these high-dimensional 
features where only the top $u$ values (ranked according to the number of
distinct source IPs involved) of each feature are
encoded for the task at hand.

We emphasize here that the feature set could be easily expanded (e.g., to include time-series
data for each Darknet scanner or encoding DNS information, when available) and the proposed
framework could easily accommodate the added features. We leave this feature expansion
as part of future work.

\begin{table*}[ht]
\centering
 \caption{Darknet features for representation learning}
 \small
 \label{tab:all_features}
\begin{tabular}{p{0.1in}|p{1in}|p{3.5in}}
\hline
ID &
  Feature &
  Description \\ 
\hline
1 &
  Total Packets   &
  Aggregate number of packets sent in the monitoring interval \\
2 &
  Total Bytes  &
  Aggregate number of bytes sent in the monitoring interval \\
3 &
  Total Lifetime   &
  Total time span of scanning activity for the scanner \\
4 &
  Number of ports scanned &
  The number of unique ports scanned by the scanner \\
5 &
  Average Lifetime &
  The average time interval that a scanner was active \\
6 &
  Average Packet Size &
  The average packet size sent by a scanner in the Darknet \\
7 &
  Set of protocols scanned  &
  One-hot-encoded set of all protocols scanned by a scanner  \\
8 &
  Set of ports scanned  &
  One-hot-encoded set of ports scanned by a scanner  \\
9 &
  Unique Destinations (Min, Max)   &
  Min and Max number of Darknet hosts scanned \\
10 &
  Unique /24 Prefixes (Min, Max)   &
  Min and Max number of Darknet /24 prefixes scanned \\
11 &
  Set of open ports at the scanner  &
  One-hot-encoded open ports/services at the scanner per \texttt{Censys.io}    \\
12 &
  Scanner's tags (e.g., device type)   &
  One-hot-encoded tags (extracted from the scanner's banner, etc.) per \texttt{Censys.io}  \\
\hline
\end{tabular}
\end{table*}

\subsection{Representation Learning}
\label{sec:deepCl}

\begin{figure}[th]
    \centering
    \includegraphics[width=\linewidth]{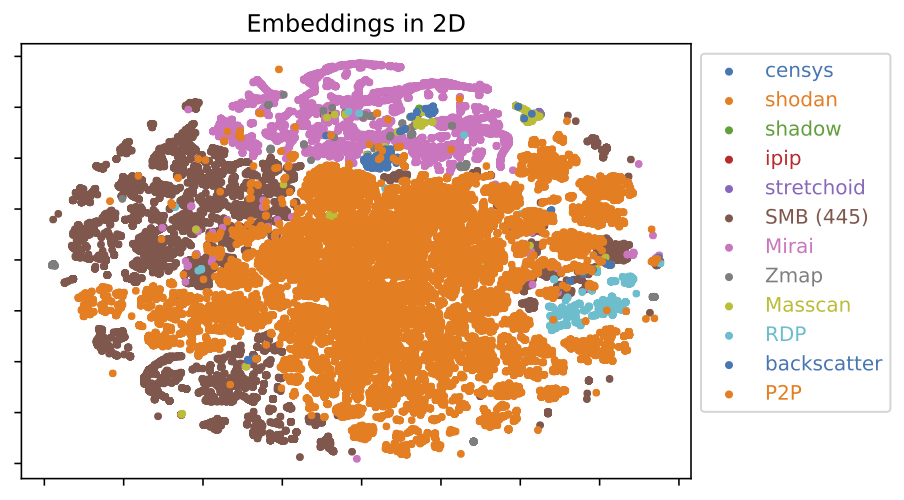}
    \caption{Embeddings in 2D showing the formation of meaningful clusters. 
    We use the t-SNE method~\cite{JMLR:v9:vandermaaten08a}
    to project in 2D the latent space of dimension $Q=50$ learned by the MLP autoencoder when applied on a set of about 330,000
    Darknet events captured in January 9th, 2021.}
    \label{fig:tsne}
\end{figure}

\begin{figure}[t]
    \centering
    \includegraphics[width=\linewidth]{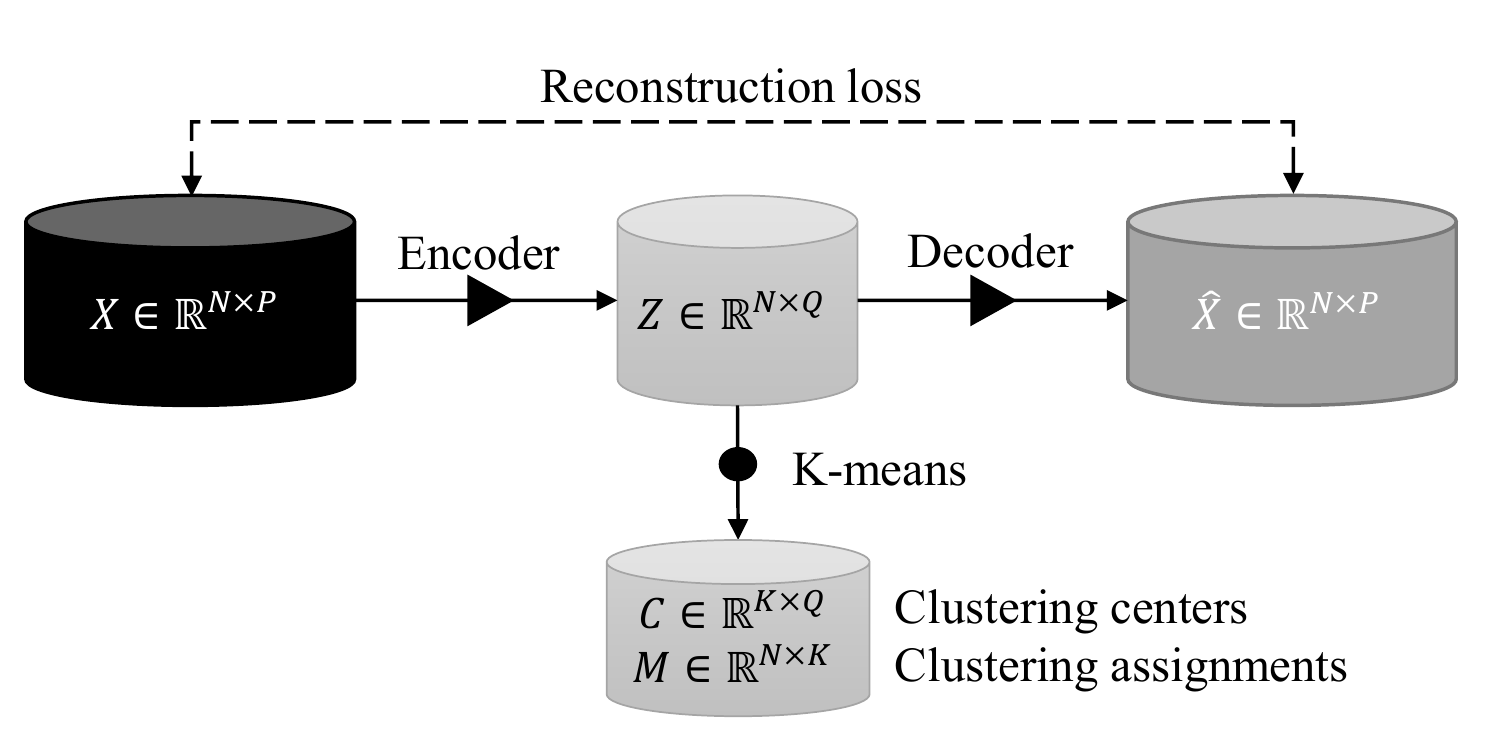}
    \caption{Autoencoding for representation learning. The embeddings
    learned are then passed to the clustering step.}
    \label{fig:deep_cluster_workflow}
\end{figure}

Motivated by the recent success of deep representation learning,
we employ the idea of autoencoders~\cite{pmlr-v70-yang17b, 10.1145/3278532.3278545, 2013arXiv1312.6114K}
to learn low-dimensional numerical embeddings of the input data. 
The resulting heterogeneity of the input data features, their high dimensionality, 
and the need to cope with potentially non-linear 
interactions between features motivated us to employ deep
representation learning to address these challenges. 
Figure~\ref{fig:tsne} illustrates how the proposed approach can
``learn" meaningful features and map them into a low-dimension latent space,
while keeping representations of similar points close together in the latent space.
These low-dimension embeddings are then passed as input to
a traditional \emph{K-means} clustering algorithm
to get the sought clusters.
The workflow is shown in Figure~\ref{fig:deep_cluster_workflow}. 

\para{Multilayer Perceptron for Autoencoding:} 
We now provide a short overview of the idea of representation learning using a Multilayer Perceptron (MLP)
architecture. 
Let $\etheta(\cdot)$ be a nonlinear encoder function parameterized by $\theta$ that maps the input data 
to a representation space, and $\dgamma(\cdot)$ be a nonlinear decoder function parameterized by $\gamma$ that maps
the data points from the representation space to the input space, such that:
\begin{align}
    &\etheta(\bx_i)=f(\bx_i;\theta)\delequal \bz_i, & f(\cdot;\theta): \mathbb{R}^P \rightarrow \mathbb{R}^Q \nonumber\\
    &\dgamma(\bz_i)=g(\bz_i;\gamma)\delequal \hat{\bx}_i, & g(\cdot;\gamma): \mathbb{R}^Q \rightarrow \mathbb{R}^P \nonumber
\end{align}

One can employ fully-connected MLP neural networks
for the implementation of both mapping functions $f(\cdot;\theta),g(\cdot;\gamma)$. In order to learn representations that preserve the information of input data, we consider minimizing the reconstruction loss:
\begin{equation}
    \min_{\theta,\gamma} \sum_{i=1}^N \left( \ell(g \circ f(\bx_i), \bx_i) \right) +  + \lambda (R(\theta)+R(\gamma))
\end{equation}
where the $\ell(\cdot): \mathbb{R}^P \rightarrow \mathbb{R}$ is a loss function that quantifies the reconstruction error. For simplicity, we choose the Euclidean distance $\ell(\bx,\bm{y})=\Norm{\bx-\bm{y}}_2^2$. 
$R(\cdot)$ is a regularization term for the model parameters
to help us avoid ``overfitting" the data.
In this work, we adopt the squared $\ell_2$ norm, such that $R(\theta)=\Norm{\theta}_2^2$. 
$\lambda \geq 0$ is the regularization coefficient.
All model parameters, \ie $\{\theta,\gamma\}$,
can be jointly learned using standard stochastic gradient-based optimization methods, 
such as Adam \cite{kingma2014adam}. 
The dimension $Q$ of the latent space, the number and size of inner layers of the MLP
architecture and other hyper-parameters (such as the regularization
coefficient) are tuned using the procedure outlined in the sequel (see Section~\ref{sec:tune}). 

\begin{figure*}
\centering
\begin{subfigure}[b]{.3\linewidth}
\includegraphics[scale=0.35]{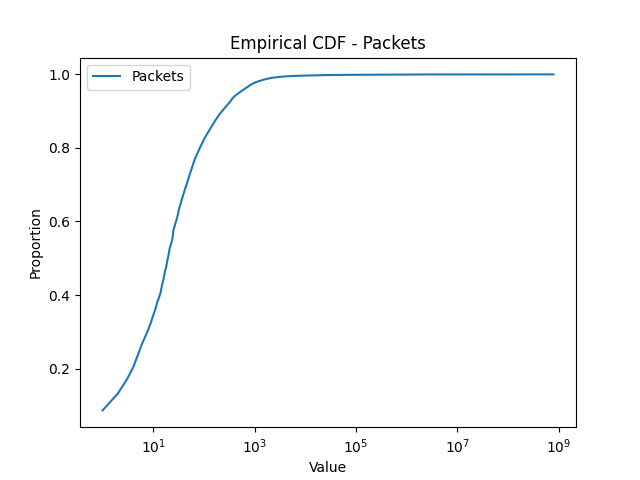}
\end{subfigure}
\begin{subfigure}[b]{.3\linewidth}
\includegraphics[scale=0.35]{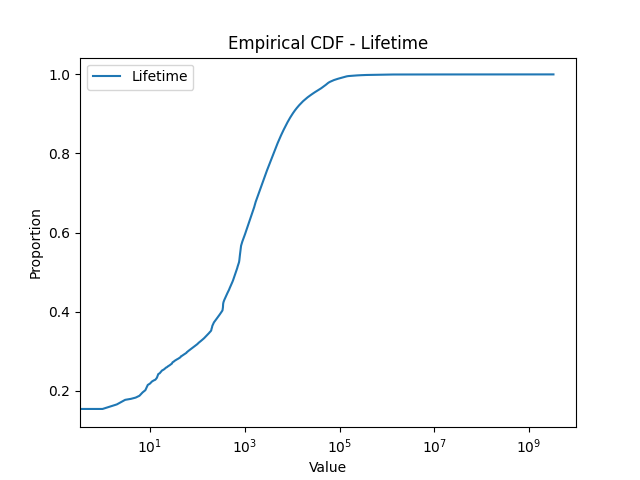}
\end{subfigure}
\begin{subfigure}[b]{.3\linewidth}
\includegraphics[scale=0.35]{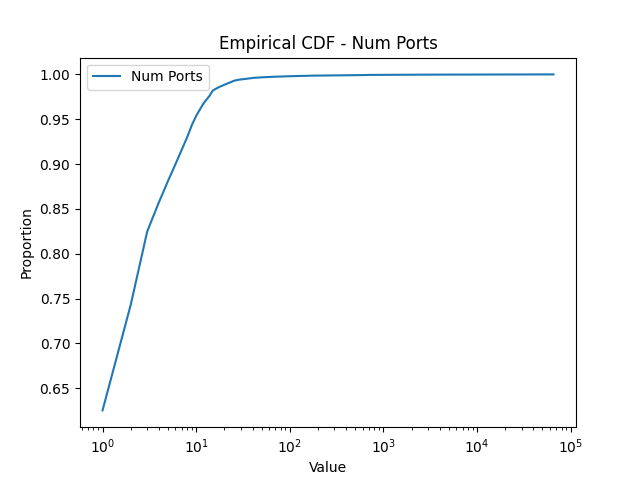}
\end{subfigure}
\begin{subfigure}[b]{.3\linewidth}
\includegraphics[scale=0.35]{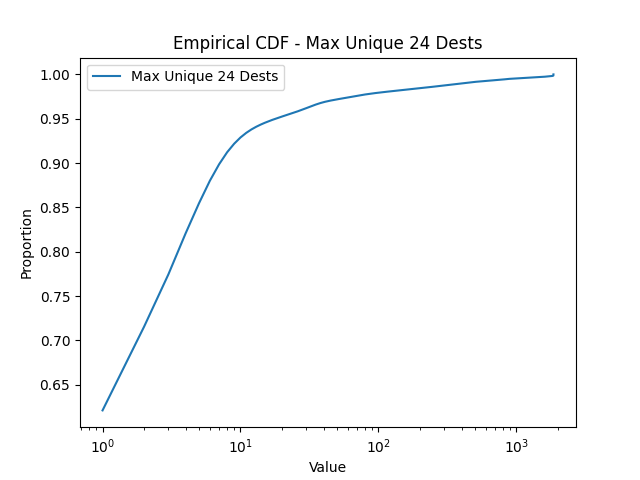}
\end{subfigure}
\begin{subfigure}[b]{.3\linewidth}
\includegraphics[scale=0.35]{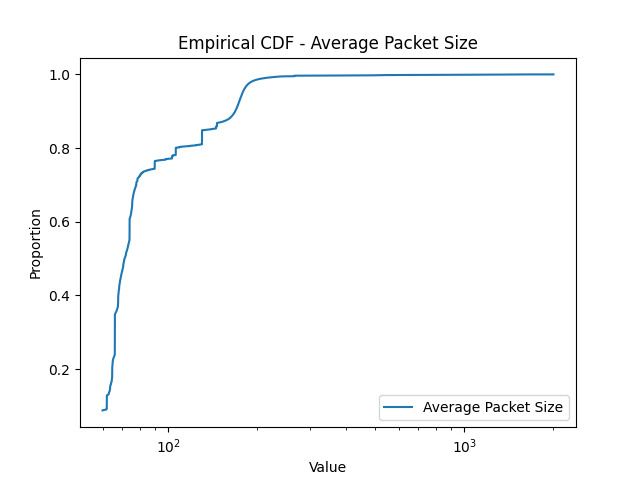}
\end{subfigure}
\begin{subfigure}[b]{.3\linewidth}
\includegraphics[scale=0.35]{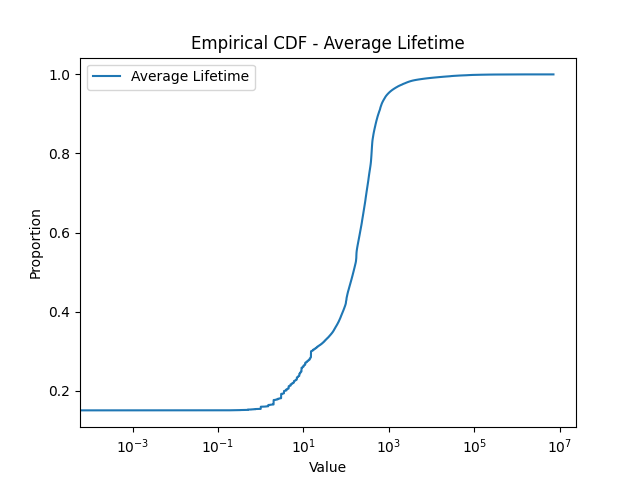}
\end{subfigure}
\caption{Distribution of numerical features. The observations are based on dataset that spans a period of 80 days.}
\label{fig:cdf}
\end{figure*}

\para{Thermometer Encoding:} One challenge associated with
encoding scanner profiles for representation learning is that a scanner profile includes, in addition to one-hot encoded binary features, numerical features (e.g., the number of ports scanned, the number of packets sent, etc). 
Mixing these two types of features might be problematic because a distance 
measure designed for one type of feature (e.g., Euclidean distance for numerical feature, 
Hamming distance for binary features) might not be suitable for the other type. To 
test this hypothesis, we also implemented an MLP network
where all (numerical) input features are encoded as \emph{binary ones} using
thermometer encoding \cite{thermometerEncoding}.
To construct the “bins” for 
the thermometer encoding, we utilize the empirical distribution of our numerical features,
shown in Figure~\ref{fig:cdf},
compiled from a dataset ranging from Nov. 1st, 2020 to Jan. 20th, 2021.
As depicted in the figure, many features, such as the one for the number of ports scanned,
exhibit a long-tail distribution.  For instance,
a very large percentage of scanners (about 70\%) scan only 1 or 2 ports, while a very small percentage of scanners scan a huge number of ports.  The latter group, while small in number, is of high interest to network analysts due to their aggressive scanning behaviors.  Therefore, we adopt log-based thermometer encoding, which enables fine-grained partition of high intensity vertical scanners.

\subsection{K-means Clustering} We perform standard \emph{K-means} clustering directly on the low-dimensional representation 
to partition the data. Formally, in this step, we aim to minimize the following clustering loss:
\begin{align}
\label{eq:kmeans}
    &\min_{C,M} \sum_{i=1}^N \ell(f(\bx_i), MC) \\
    &\mathrm{s.t.}~m_{i,j}\in \{0,1\},~ M 1_{K} = 1_{N} \nonumber
\end{align}
where $M$ is the clustering assignment matrix, the entries of which are all binary. $C$ is the matrix of clustering centers that lie in the representation space. $1_{K}$ is a $K$-dimensional column vector of ones. 
We utilized broadly available libraries for the implementation
of the representation learning and clustering steps, such as \texttt{PyTorch}
and \texttt{scikit-learn}.


\section{Evaluation of Clustering and Representation Learning}
\label{sec:peava}

\subsection{Clustering Quality Measures}

To evaluate the quality of the clustering results, 
we utilized the \emph{silhouette score}, which is the standard measure for clustering quality,
along with two additional \emph{evaluation metrics} to obtain a more
comprehensive assessment: 
the \emph{Jaccard score}~\cite{halkidi2001survey} that is based on 
a set of external labels (created for evaluation purpose, 
and not used by clustering) based on domain knowledge (e.g., known
fingerprints of Mirai malware, known port vulnerabilities
such as port 445 associated with the attacks on the SMB protocol, 
known heavy scanners such as \texttt{shodan.io}, \texttt{Censys.io}, etc.) 
to determine the similarity of the clustering
outcomes with the partition formed by the labels; 
and a \emph{stability score} that evaluates the stability of clustering results by comparing the similarity of clusters generated by different sub-samples.

\para{Silhouette Coefficient:} The silhouette coefficient is frequently used for
assessing the performance of unsupervised clustering algorithms~\cite{Rousseeuw87Silhouettes}. Clustering
outcomes with ``well defined" clusters (i.e., clusters that are tight and 
well-separated from peer clusters) get a higher silhouette coefficient score. Formally,
the silhouette coefficient is obtained as
\begin{equation}
    \mathit{sc} = \frac{b-a}{\max\{a, b\}},
\end{equation}
where $a$ is the average distance between a sample and all the other points in the same cluster and
$b$ is the average distance between a sample and all points in the next nearest cluster.

\para{Jaccard Score:} The Jaccard index or Jaccard similarity coefficient is a commonly used distance metric to assess the similarity of two finite sets. It measures this similarity as the ratio of intersection and union of the sets. This metric is, thus, suitable for quantitative evaluation of the clustering outcomes. Given that there is a domain inspired predefined partitioning $P = \{P_1,P_2,\ldots,P_S\}$ of the data, the distance or the Jaccard Score of the clustering result $C = \{C_1,C_2,\ldots,C_N\}$ on the same data is computed as~\cite{halkidi2001survey}:

\begin{equation}
    \mathit{Jaccard} = \frac{M_{11}}{M_{01} + M_{10} + M_{11}},
\end{equation}

where $M_{11}$ is the total number of pair of points that belong to the same group in $C$ as well as the same group in $P$, $M_{01}$ is the total number of pair of points that belong to the different groups in $C$ but to same group in $P$ and $M_{10}$ is the total number of pair of points that belong to the same group in $C$ but to different groups in $P$.

\begin{figure}[ht]
    \centering
    \includegraphics[width=3in]{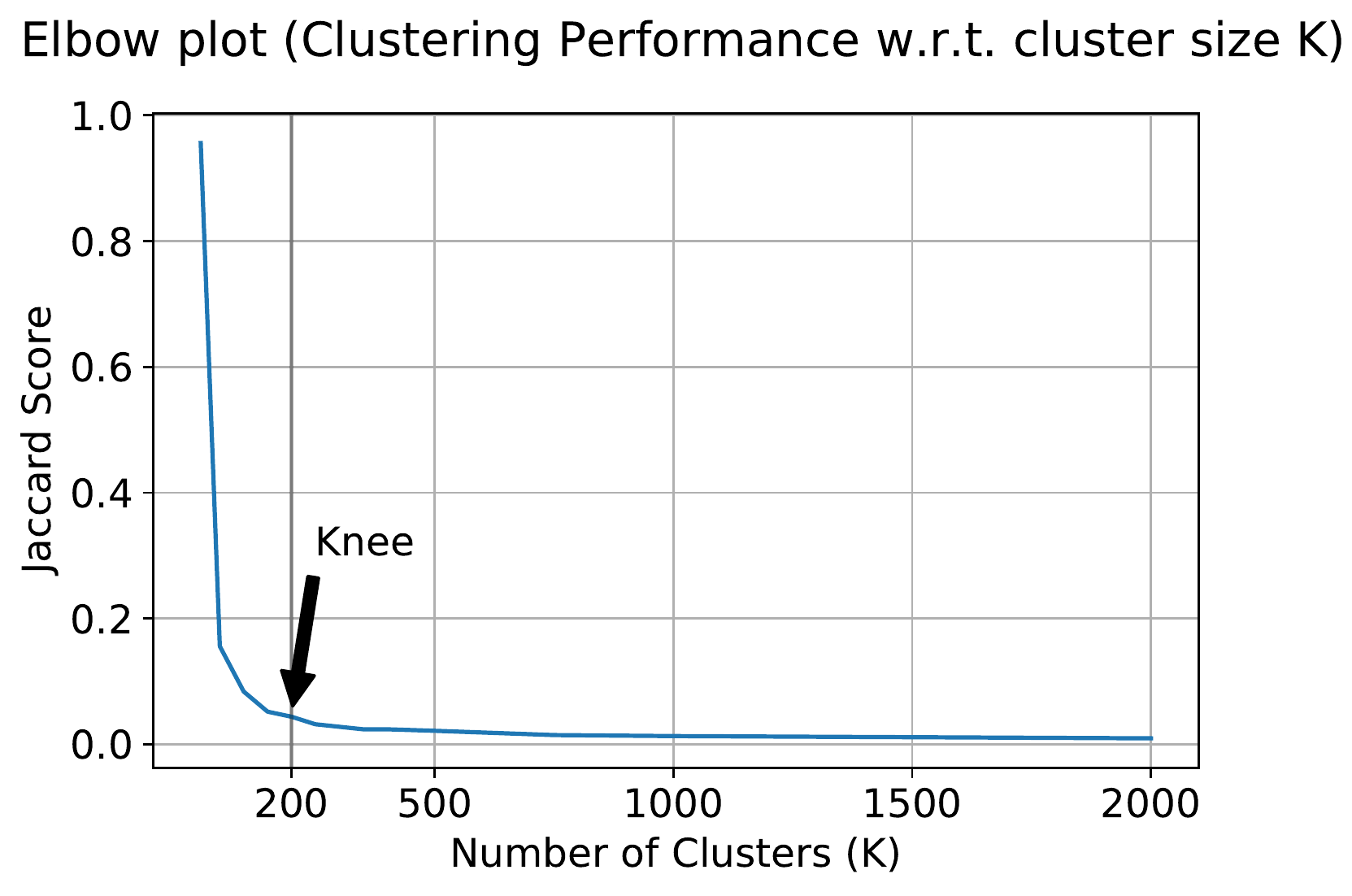}
    \caption{This elbow plot shows the Jaccard scores obtained from clustering the dataset into different number of clusters ($K$). The Jaccard score, which starts at a higher value for lower values of $K$, decreases rapidly at the beginning and after certain value of $K$, the decrease becomes slower. This value of $K(=200)$ marks the ``knee"
    of the plot and provides a good choice for the number of clusters to use in the clustering step.}
    \label{fig:sizeK}
\end{figure}

This cluster evaluation metric incorporates domain knowledge (such as
Mirai, Zmap and Masscan scanners, that can identified
by their representative packet header signatures~\cite{MiraiUSENIX2017, Durumeric:2014:IVI:2671225.2671230}, and other
partitions as outlined earlier)
and measures how compliant the clustering results are with the known partitions. Jaccard score decreases as we increase the number of clusters used for clustering. This decrease is drastic at the beginning and slows down eventually forming a ``knee" (see Figure~\ref{fig:sizeK}). The ``knee" where the significant local change in the metric occurs reveals the underlying number of groupings in the data \cite{halkidi2001survey}.  

\para{Cluster Stability Score:} 
Quantifying cluster stability is important because it assesses how
clustering results vary due to different sub sampling from the data.
A clustering result that is not sensitive to sub-sampling, hence more stable,
is certainly more desirable.  
In other words, the cluster structure uncovered by the clustering
algorithm should be similar across different samples from the same data distribution \cite{luxburg2010stability}.

In order to analyze the stability of the clusters, we generate multiple subsampling versions of the data by using bootstrap resampling. These samples are clustered individually using the same clustering algorithm. The cluster stability score is, then, the average of the pairwise distances between the clustering outcomes of two different subsamples. For each cluster from one bootstrap sample, we identify its most similar cluster among clusters from another bootstrap sample using Jaccard index as the pairwise distance metric. In this case, the Jaccard index is simply the ratio of the intersection and union between the clusters. The average of these Jaccard scores across all pairs
of samples provides a measure of how stable the clustering results are. 

\subsection{Hyper-parameter Tuning}
\label{sec:tune}

Systematic tuning of the MLP hyper-parameters is critical
to ensure that the network learns an information-preserving low dimension 
representation of the high dimensional input data.
Following best practices for hyper-parameter tuning in machine learning,
the set of hyper-parameters is adjusted to find parameter values
that result in low \emph{autoencoder loss} based on evaluation data
not used in the training phase of the autoencoder. Hyper-parameters tuned include
the size $Q$ of the embeddings (i.e., latent space),
the regularization coefficient,
the ``learning rate" used in stochastic gradient descent, 
the ``dropout" probability employed to prevent over-fitting the training data,
the number of ``epochs"
and the ``batch size". 

We employed the \texttt{Ray[Tune]} framework~\cite{raytune} to tackle the important task of hyper-parameter tuning.
Based on the tuning results, we choose the dimension of the latent space to be 50. 
The “best model” for both MLP and thermometer encoding MLP 
have a hidden layer 
of dimension 1000.
Using 100 epochs, a batch size of 2000 data points
and a value of 0.001 for learning rate, dropout probability and regularization weight
provided optimal performance. 
All results presented in the following sections were generated using these 
tuned hyper-parameters settings.

\subsection{Evaluating Clustering Results}

\begin{table*}[t]
    \centering
    \caption{Evaluation of Clustering through Representation Learning}
    \label{table:Eval-Clustering-RepLearn}
    \begin{tabular}{c|c|c|c|c}
    \hline
    \textbf{Autoencoder}   & \textbf{Loss} & \textbf{Silhouette} & \textbf{Jaccard} & \textbf{Stability} \\
     \hline 
      MLP   & 0.96 (1.71) & 0.44 (0.01) & 0.043 (0.001) & 0.40 (0.007) \\
      Thermo-MLP  & 24.97 (1.44) & 0.58 (0.02) & 0.012 (0.001) & 0.51 (0.008) \\
      \hline
    \end{tabular}
\end{table*}

To evaluate the clustering outcomes of the two autoencoder architectures we
utilize the 3 metrics outlined earlier.
We employ a Darknet dataset compiled for the day of January 9th, 2021, which includes about 2 million scanners. 
Figure ~\ref{fig:sizeK} shows the clustering performance for different
number of clusters ($K$).  
Following the suggestions in \cite{halkidi2001survey}, we thus
select the number of clusters for clustering Darknet data to be 200.
A random sample of 500K scanners were used to perform 50 iterations of training autoencoders and k-means clustering, using 50K scanners in each iteration.  The mean and standard deviation of the three clustering evaluation metrics, as well as the mean and standard deviation of the loss function (L2 for MLP, Hamming distance for thermometer-encoding-based MLP (TMLP)), are shown in Table~\ref{table:Eval-Clustering-RepLearn}.

The results indicated that the TMLP autoencoder led to better clustering results based on the silhouette and stability scores.  
However, a smaller Jaccard score was reported when compared to the MLP autoencoder.
By inspecting the clusters generated, we  noticed 
that this is probably due to the fact that
TMLP tended to group scanners into smaller clusters that are similar in size.  
I.e., it generated multiple fine-grained clusters that correspond to a common large 
external label used for external validity measure (i.e., the Jaccard score).  Because our current Jaccard score 
computation does not take into account the hierarchical structure of external label, 
fine-grained partition of external labels are penalized, even though they can 
provide valuable characteristics of subgroups in a malware family (e.g., Mirai).
Henceforth, though, we present results using the MLP architecture that
scored very well on all metrics and 
provided more interpretable results.

\section{Interpretation and Internal Structure of Clusters}
\label{sec:odt}

Clustering interpretation is critical in explaining the clustering outcome
to network analysts. Contrary to supervised learning tasks, there is no ``correct"
clustering assignment and the clustering outcome is a consequence of the features employed.
Hence, it is germane to provide interpretable and simple rules that explain the clustering outcome
to network analysts so that they are able 
to i) compare clusters and assess inter-cluster similarity,
ii) understand what features (and values thereof) are responsible for the formation of a given cluster,
and iii) examine the hierarchical relationship amongst the groups formed.

\subsection{Optimal Decision Trees}

We propose the use of \emph{decision trees}~\cite{hastie01statisticallearning} 
for clustering interpretation.
Decision trees are conceptually simple, yet powerful, 
for supervised learning tasks (i.e., when labels are available)
and their simplicity makes them easily understandable by human analysts. 
Specifically, we are interested in \emph{classification trees}.

In a classification tree setting, one is given $N$ observations that
consist of $p$ inputs, that is $x_i = (x_{i_1}, x_{i_2}, \ldots, x_{i_p})$,
and a target variable $y_i$. The objective is to recursively 
partition the input space and assign the $N$ observations
into a classification outcome taking values $1, 2, \ldots, K$ such that
the classification error is minimized. For our application, the $N$
observations correspond to the $N$ Darknet events we had clustered and the 
$K$ labels correspond to the labels assigned by the clustering step. 
The $p$ input features are closely associated with the $P$ features used in the
representation learning step (see Section~\ref{sec:clusteringScanners}). Specifically, we still employ all the numerical features
but we also introduce the new binary variables / tags shown in Table~\ref{table:orion:tags}. These
``groupings", based on domain knowledge, succinctly summarize some notable Darknet activities we are aware of (e.g., Mirai
scanning, backscatter activities, etc.) and, we believe can help the analyst easily interpret the
decision tree outcome.

Traditionally, classification trees are constructed using heuristics to split the input
space~\cite{Breiman1983ClassificationAR, bertsimas17optimal, hastie01statisticallearning}. 
These greedy heuristics though lead to trees that are ``brittle", i.e., trees that can
drastically change even with the slightest modification in the input space
and therefore do not generalize well. One can overcome this by using tree
ensembles or ``random forests" but this option is not suitable for the interpretation task at hand since
one then needs to deal with multiple trees to interpret a clustering outcome. Hence,
we decided to work with \emph{optimal classification trees}~\cite{bertsimas17optimal}
which are nowadays feasible to construct due to recent algorithmic 
advances in mixed-integer optimization and hardware improvements that speed-up computations. 

\begin{figure}[ht]
    \centering
    \includegraphics[width=3in,trim=0 1cm 0 1cm,clip]{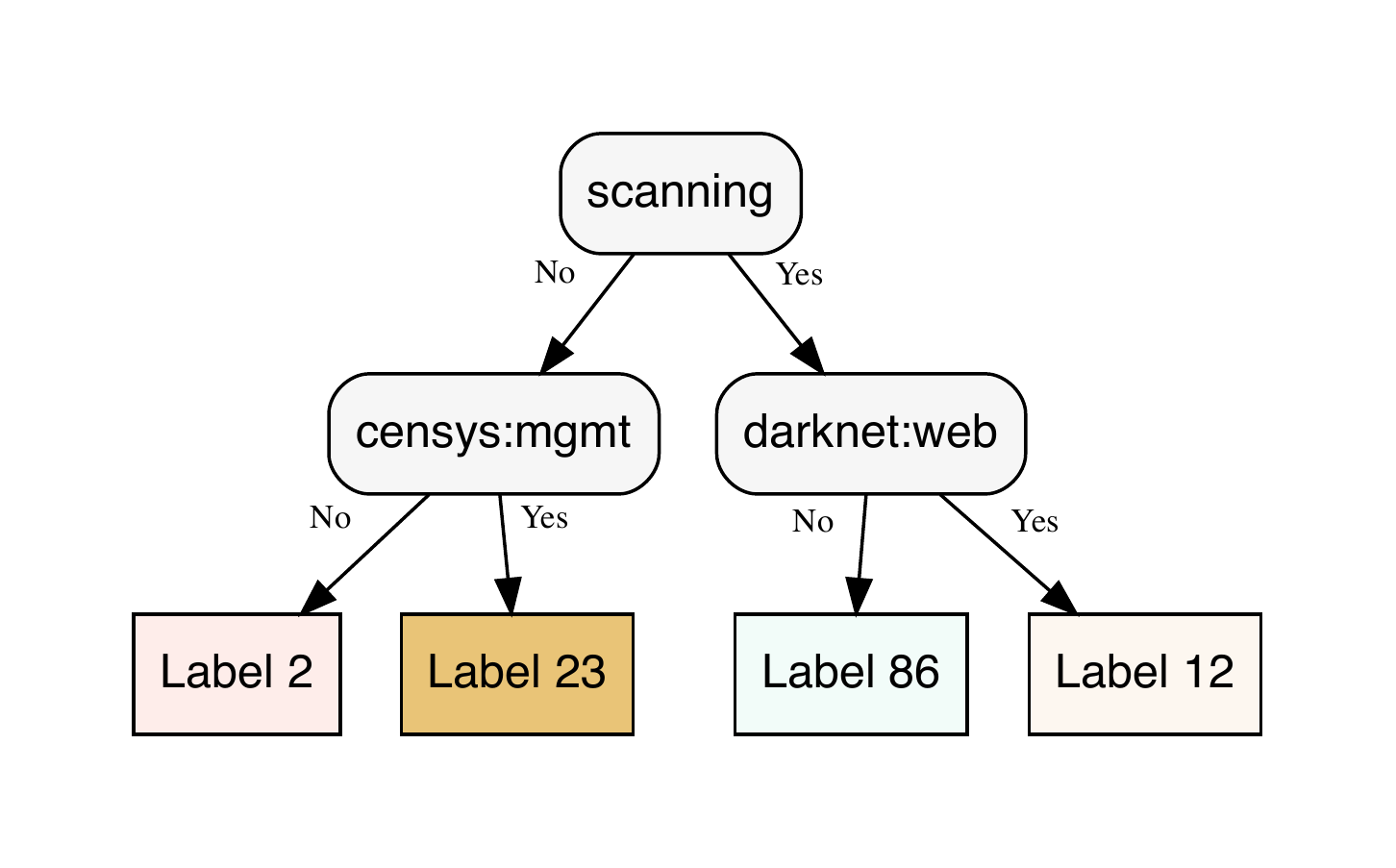}
    \caption{Optimal classification tree of depth 3.}
    \label{fig:toytree}
\end{figure}

\begin{figure}[ht]
    \centering
    \includegraphics[width=3in,trim=0 1cm 0  1cm,clip]{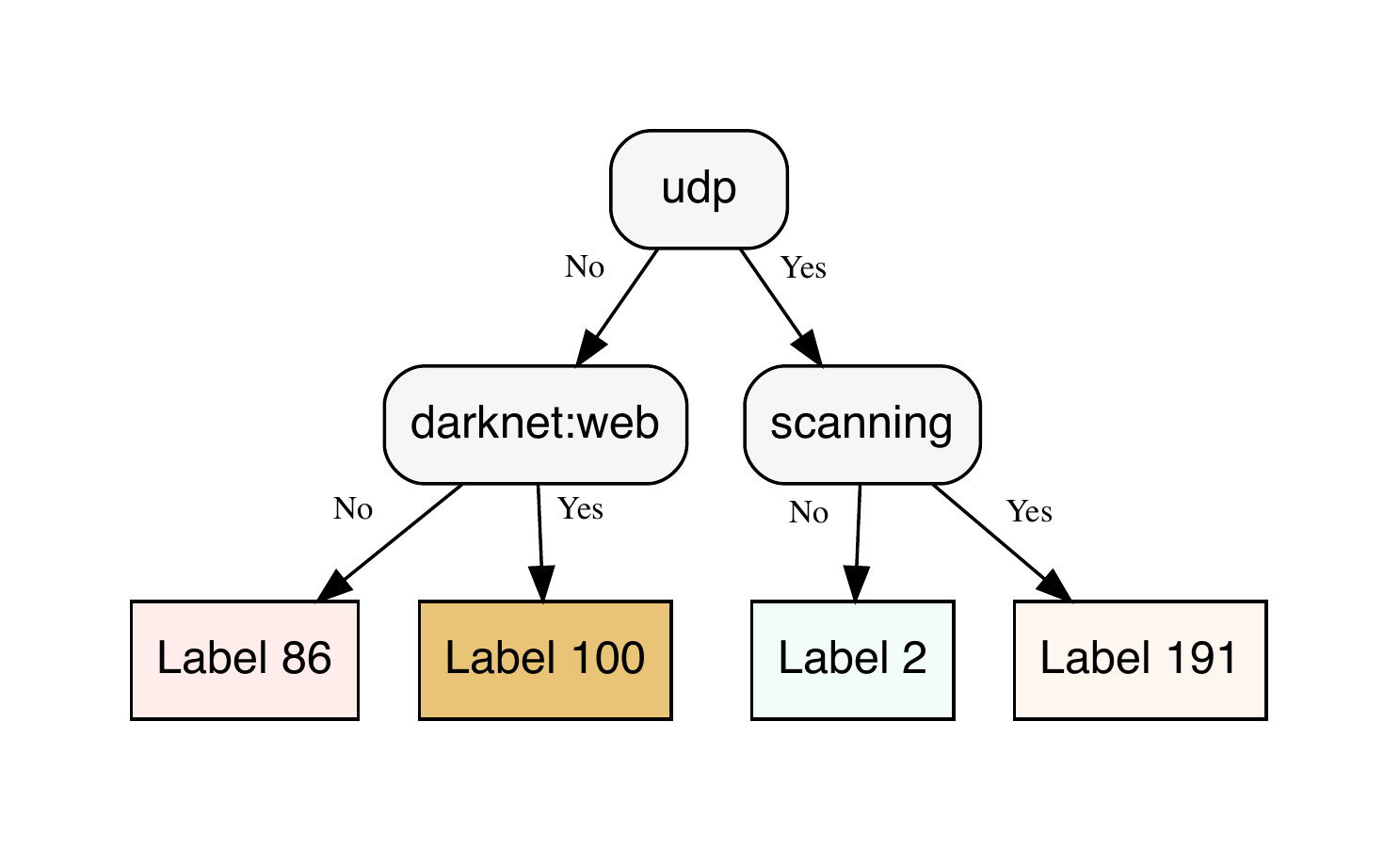}
    \caption{Non-optimal classification tree of depth 3.}
    \label{fig:toytree-nonoptimal}
\end{figure}

We used the software provided by \texttt{interpretable.ai}~\cite{interpretableai} to generate all optimal decision 
trees shown in the sequel. Figure~\ref{fig:toytree}
shows an example generated for 467,293 Darknet events for Sept.\ 14th, 2020. 
The structure of the tree, albeit minimal, is quire revealing.
First, the leaves correspond to the largest 4 clusters (with sizes
14953, 11013, 10643 and 9422, respectively) 
found for Sept.\ 14th which means that the clusters with the most impact are captured. 
Another important observation is that the type of decision rules used to split
the input space (namely, \emph{scanning}, \emph{censys:mgtm} and \emph{orion:remote})
are indicative of the main Darknet activities during that day. 
Comparing with the non-optimal, heuristic-based decision tree of Figure~\ref{fig:toytree-nonoptimal},
we recognize some important differences: 1) two new clusters have emerged (with labels 100 and 191)
that \emph{do not} rank within the top-4 clusters (they rank 8th and 10th, respectively, with 6977 and
6404 members); and 2) there is some ``redundancy" 
in the decision rules used for splitting when both the tags UDP and ``scanning"
are present. This is because UDP and scanning (i.e., TCP SYN requests and ICMP Echo Requests)
are usually complementary to each other. 

\begin{table}[t]
\small
\centering
\caption{Groupings Used for Clustering Interpretation.}
\label{table:orion:tags}
\begin{tabular}{l|l}

\hline
\textbf{Feature}    & \textbf{Description}                                                         \\ \hline
darknet:web           & Ports: 80, 443, 81                                                           \\ 
darknet:remote        & Ports: 22, 23                                                                \\ 
darknet:mssql         & Ports: 1433                                                                  \\ 
darknet:samba         & Ports: 445                                                                   \\ 
darknet:rdp           & Ports matching regex '\textbackslash{}d+3389\textbackslash{}d+'              \\ 
darknet:quote         & Port: 17                                                                     \\ 
darknet:p2p           & Ports matching regex '17\textbackslash{}d\textbackslash{}d\textbackslash{}d' \\ 
darknet:amplification & Ports: 123, 53, 161, 137, 1900, 19, 27960, 27015                             \\ 
censys:web           & Tags: http, https                                                            \\ 
censys:remote        & Tags: ssh, telnet, remote                                                    \\ 
censys:mssql         & Tags: mssql                                                                  \\ 
censys: samba        & Tags: smb                                                                    \\ 
censys:embedded      & Tags: embedded, DSL, model, iot                                              \\ 
censys:mgmt          & Tags: cwmp, snmp                                                             \\ 
censys:storage       & Tags: ftp, nas                                                               \\ 
censys:amplification & Tags: dns, ntp, memcache                                                     \\ 
scanning             & TCP and ICMP scanning                                                       \\ 
backscatter          & Protocols / flags associated with backscatter                                \\ 
UDP                  & Whether its UDP                                                              \\ 
Unknown / other      & Other protocols / flags                                                      \\\hline 
\end{tabular}
\end{table}

\subsection{Internal Structures of Clusters}

One of the important challenges in clustering is identifying characteristics 
of a cluster that distinguish it from other clusters.  While the center of a cluster 
is one useful way to represent a cluster, it can not clearly reveal the features and 
values that define the cluster.  This is even more challenging for characterizing clusters 
of high-dimensional data, such as the scanner profiles in the Darknet.
One can address this challenge by defining “internal structures”  
based on the decision trees learned.  

Given a set of clusters ${C_1, C_2, \ldots, C_K}$ that form
a partition of a dataset $D$, a disjunctive normal forms (DNF) $S_i$ is said 
to be an internal structure of cluster $C_i$ if any data items in D satisfying 
$Si$ are more likely to be in $C_i$ than in any other clusters.  
Hence, an internal structure of a cluster captures characteristics of the cluster that distinguishes it from all other clusters.  
More specifically, the conjunctive conditions of a path in the decision tree to a leaf node that predicts cluster $C_i$ forms the conjunctive (AND) component of the internal
structure of $C_i$.  Conjunctive path description from multiple paths in the decision tree that predict the 
same cluster (say $C_i$) are combined into a disjunctive normal form that characterizes the cluster 
$C_i$.  
Hence, the DNF forms revealed by decision tree learning 
on a set of clusters expose the internal structures of these clusters. 
We plan to pursue detection of structural (di)similarities in the Darknet based on DNF forms applied
on decision tree outcomes as part of future work.

\section{Detecting Cluster Changes}
\label{sec:emd}

Given the proposed clustering framework, one can readily obtain Darknet clusters
on a daily basis (or at any other granularity of interest) and \emph{compare} the
clustering outcomes to glean insights on their similarities. This is critical
to security analysts aiming to automatically track changes in the behavior of the Darknet,
in order to detect new emerging threats or vulnerabilities in a timely manner.

For example, Figure~\ref{fig:emd} (left panel) tracks the evolution of the Darknet for the
whole month of September 2020. We compare the clustering outcome of \emph{consecutive} days 
using a distance metric applied on the clustering profile of each pair of days. 
Specifically, we
use the \emph{Earth Mover's Distance}~\cite{710701} which is a measure 
that captures the dissimilarity between 
two multi-dimensional distributions (also known as Wasserstein metric).  
Intuitively, by considering
the two distributions as two piles of dirt spread in space,
the Earth Mover's Distance captures the minimum cost required
to transform one pile to the other. The cost here is defined
as the distance (Euclidean or other appropriate distance) travelled to transfer a unit amount of dirt
times the amount of dirt transferred. This problem can be formulated as a linear 
optimization problem and several solvers are readily available (e.g.,~\cite{pywasserstein}).

In our setting, each clustering outcome defines a distribution or ``signature"
that can be utilized for comparisons. Specifically, denote the set of clusters
obtained after the clustering step as $\{C_1, C_2, \ldots, C_K\}$ and the centers
of all clusters as $\{m_1, m_2, \ldots, m_K\}$ where 
$$
m_i = \frac{\sum_{j \in C_i} x_{j}}{|C_i|},  
$$
$i = 1,\ldots, K$, and $x_j \in \mathbb{R}^P, j = 1,\ldots, N$.
Then, the signature
$$
S = \{(m_1, w_1), (m_2, w_2), \ldots, (m_K, w_K)\}
$$
can be employed, where $w_i$ represents the ``weight`` of cluster $i$
which is equal to the fraction of items in that cluster over the total
population of scanners. The results we present in the next section
were compiled by applying this signature on the clustering outcome of each day;
research on other signatures (possibly of lower dimensionality) is part of ongoing work. 

\section{Analysis of Real-world Events}
\label{sec:real}

\begin{figure*}[ht]
    \centering
    \includegraphics[width=\linewidth]{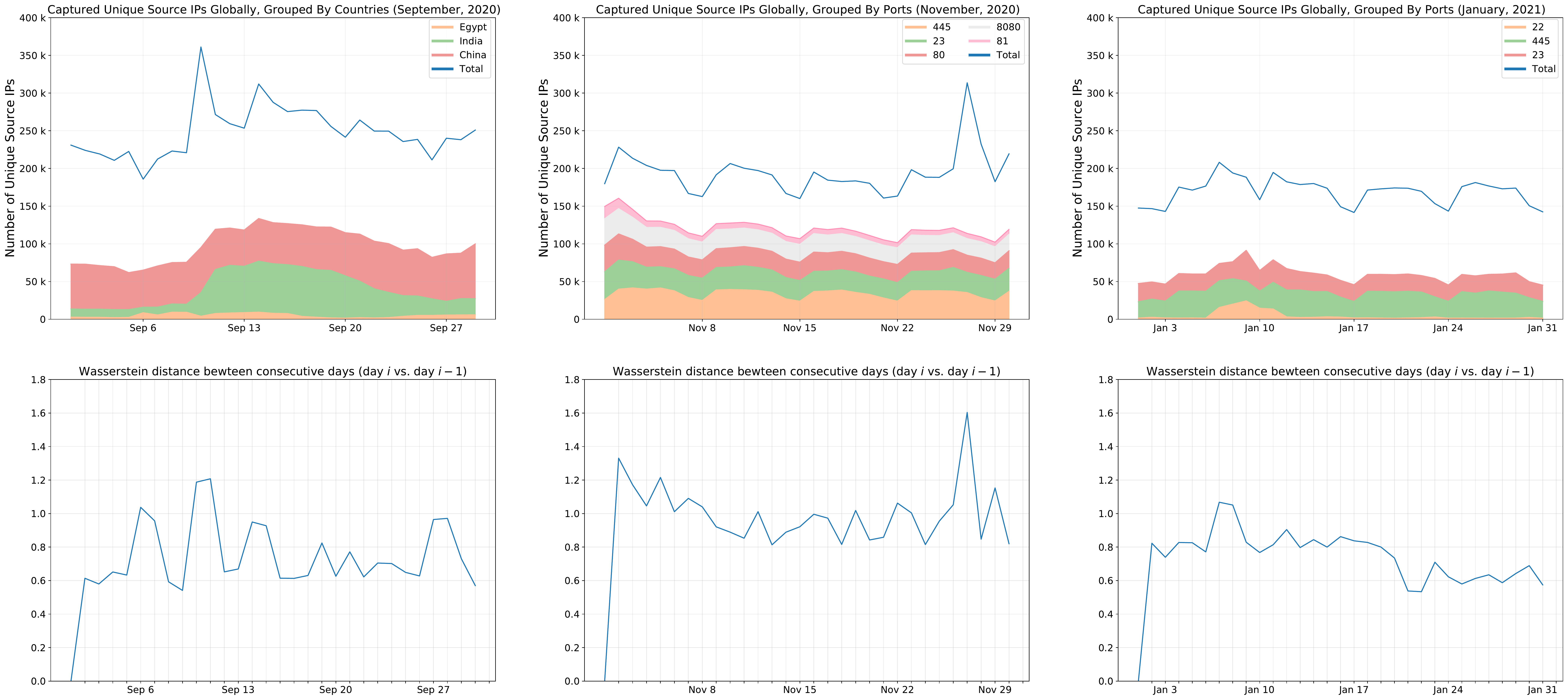}
    \caption{Diagnosing temporal clustering changes using Earth Mover's Distance.}
    \label{fig:emd}
\end{figure*}

We are now ready to demonstrate the entire methodology 
(i.e., clustering, detection of longitudinal structural changes using Earth Mover's Distance (EMD),
and decision trees for interpretation) when applied to three months of
data from a large operational /13 Darknet. The objective 
of this section is to showcase that 1) important Internet-wide
events were discovered using the proposed novelty detection
approach (see Figure~\ref{fig:emd}) and 2) to
highlight the importance of clustering interpretation using
decision trees. The case studies considered are summarized below:
\begin{itemize}
    \item September 2020: A large Mirai outbreak that emerged on September 6th, 2020, primarily originating from Egypt and India;
    \item November 2020: A November 27th spike in the number of infected hosts appearing to scan our Darknet, attributed primarily to embedded devices;
    \item January 2021: The emergence of a large number of infected CWMP-enabled devices, associated with a tier-1 US ISP, involved in aggressive SSH scanning activities.
\end{itemize}

To underline the importance of integrating external data sources in understanding Darknet data,
all results presented henceforth are for Darknet scanners / IPs with available Censys-based features
(see Table~\ref{tab:all_features} for the list of our features).

\subsection{The Mirai incident}

Figure~\ref{fig:emd} (bottom row, left plot) shows a significant increase in the EMD distance between the clustering
outcome of Sept.\ 5th and Sept.\ 6th. Figure~\ref{fig:emd} (top row, left plot) shows the aggregate number of unique scanners
for the total month of September and a detailed view of the top-3 countries for that month (regarding unique scanners).
With a bare eye we observe an increase in scanners originating from Egypt, but looking at a mere scanning volume
cannot let us diagnose what really happened. We therefore look at the top clusters extracted for both days
for an answer.

\begin{table}[htb]
    \centering
    \caption{Top-10 clusters for Sept.~5, 2020}
    \footnotesize
    \label{table:top_sept5}
    \resizebox{\linewidth}{!}{\begin{tabular}{| c |c || c | c || c | c || c |}
    \hline
    \textbf{\thead{Cluster\\ Label}}   & \textbf{\thead{Cluster\\ Size}} & \textbf{\thead{Top\\ Ports}}   & \textbf{\thead{Top Port\\ Fraction}} & \textbf{\thead{Top\\ Tags}} & \textbf{\thead{Top Tag\\ Fraction}} &
    \textbf{\thead{Mirai\\ Coverage}}\\
     \hline 
      73 & 9587        &   54148    & 0.36 & telnet & 0.26 & 0.00\\
      3 & 8513        &   445    & 0.73 & ftp & 0.44 & 0.00\\
      57 & 8053        &   54148    & 0.77 & cwmp & 0.94 & 0.00\\
      21 & 7293        &   8080    & 0.32 & cwmp & 0.94 & 0.00\\
      188 & 5817        &   8080    & 0.28 & ftp & 0.46 & 0.00\\
      87 & 5052        &   445    & 0.65 & http & 0.87 & 0.00\\
      171 & 5024        &   17130    & 0.13 & cwmp & 0.96 & 0.00\\
      106 & 4818        &   17140    & 0.50 & dns & 0.83 & 0.00\\
      135 & 4602        &   23    & 0.27 & dns & 0.80 & 0.00\\
      26 & 4550        &   54148    & 0.59 & http & 0.77 & 0.00\\\hline
      
    \end{tabular}}
\end{table}

\begin{table}[htb]
    \centering
    \caption{Top-10 clusters for Sept.~6, 2020}
    \footnotesize
    \label{table:top_sept6}
    \resizebox{\linewidth}{!}{\begin{tabular}{| c |c || c | c || c | c || c |}
    \hline
    \textbf{\thead{Cluster\\ Label}}   & \textbf{\thead{Cluster\\ Size}} & \textbf{\thead{Top\\ Ports}}   & \textbf{\thead{Top Port\\ Fraction}} & \textbf{\thead{Top\\ Tags}} & \textbf{\thead{Top Tag\\ Fraction}} &
    \textbf{\thead{Mirai\\ Coverage}}\\
     \hline 
      6 & 15457        &   52695    & 0.23 & smb & 0.18 & 0.00\\
      49 & 11612        &   445    & 0.88 & smb & 0.29 & 0.00\\
      17 & 9759        &   52695    & 0.46 & cwmp & 0.94 & 0.00\\
      2 & 9744        &   0    & 0.98 & dns & 0.89 & 0.00\\
      7 & 7335        &   23    & 0.92 & ftp & 0.40 & 0.88\\
      47 & 6225        &   80    & 0.34 & cwmp & 0.94 & 0.00\\
      53 & 6191        &   8080    & 0.15 & snmp & 0.42 & 0.01\\
      38 & 6163        &   52695    & 0.05 & dns & 0.98 & 0.00\\
      78 & 6065        &   8080    & 0.30 & snmp & 0.56 & 0.00\\
      83 & 5311        &   0    & 0.91 & dns & 0.81 & 0.00\\\hline
      
    \end{tabular}}
\end{table}

Tables~\ref{table:top_sept5} and~\ref{table:top_sept6} tabulate the top-10 clusters for each day.
Notice that the column ``Mirai Coverage" (which denotes the portion of Darknet scanners within that 
cluster that include the well-known Mirai fingerprint~\cite{MiraiUSENIX2017} in the 
packet header of the packets they emitted) is zero
for all top clusters of Sept.\ 5th but a relatively large Mirai-related cluster emerges on Sept.\ 6th.
Clearly, we now have a good indication about what might have caused the shift in the clustering
structure between Sept.\ 5th and 6th.

Looking at our data more closely we found that 
there were 15970 total Mirai victims on Sept.\ 5 and 28261 on Sept.\ 6th.
Table~\ref{table:top_mirai_Sept5} shows the top Mirai-related clusters for Sept.\ 5th
and Table~\ref{table:top_mirai_Sept6} the top ones for the next day. As observed,
the Mirai clusters have increased significantly. Notice also
that Censys can provide some valuable information for interpreting each cluster;
the largest Mirai cluster on Sept.\ 6th seems to be heavily associated with devices
having the FTP port open.

Examining the distribution of countries affected by the Mirai malware (see Table~\ref{table:top_country},
recall that we geo-annotate all scanners for result interpretation, but we
don't use this information as a clustering feature), we see Egypt emerging as the number one affected country.
A few days later, India will be facing a similar Mirai outbreak (the large
increase in unique scanners seen in Figure~\ref{fig:emd} for Sept.\ 10th
is India's Mirai outbreak with scanners having a similar profile like the ones observed 
in Egypt).

\begin{table}[htb]
    \centering
    \caption{Top 5 Mirai-related Clusters on Sept.~5, 2020}
    \footnotesize
    \label{table:top_mirai_Sept5}
    \resizebox{\linewidth}{!}{\begin{tabular}{| c |c || c | c || c | c || c |}
    \hline
    \textbf{\thead{Cluster\\ Label}}   & \textbf{\thead{Cluster\\ Size}} & \textbf{\thead{Top\\ Ports}}   & \textbf{\thead{Top Port\\ Fraction}} & \textbf{\thead{Top\\ Tags}} & \textbf{\thead{Top Tag\\ Fraction}} &
    \textbf{\thead{Mirai\\ Coverage}}\\
     \hline 
      147 & 3183        &   2323    & 0.05 & http & 0.64 & 0.95\\
      55 & 252        &   23    & 0.45 & http & 0.50 & 0.94\\
      85 & 780        &   5555    & 0.53 & http & 0.73 & 0.92\\
      92 & 1192        &   23    & 0.03 & http & 0.61 & 0.91\\
      159 & 534        &   23    & 0.15 & http & 0.53 & 0.90\\\hline
      
    \end{tabular}}
\end{table}

\begin{table}[htb]
    \centering
    \caption{Top 5 Mirai-related Clusters on Sept.~6, 2020}
    \footnotesize
    \label{table:top_mirai_Sept6}
    \resizebox{\linewidth}{!}{\begin{tabular}{| c |c || c | c || c | c || c |}
    \hline
    \textbf{\thead{Cluster\\ Label}}   & \textbf{\thead{Cluster\\ Size}} & \textbf{\thead{Top\\ Ports}}   & \textbf{\thead{Top Port\\ Fraction}} & \textbf{\thead{Top\\ Tags}} & \textbf{\thead{Top Tag\\ Fraction}} &
    \textbf{\thead{Mirai\\ Coverage}}\\
     \hline 
      15 & 538        &   2323    & 0.76 & http & 0.52 & 0.98\\
      103 & 3580        &   2323    & 0.02 & http & 0.76 & 0.93\\
      178 & 2920        &   23    & 0.71 & http & 0.64 & 0.91\\
      7 & 7335        &   23    & 0.92 & ftp & 0.40 & 0.89\\
      40 & 1135        &   23    & 0.02 & http & 0.60 & 0.88\\\hline
      
    \end{tabular}}
\end{table}

\begin{table}[ht]
\centering
\caption{Distribution of Mirai By country}
\label{table:top_country}
\subfloat[][Sept.\ 5, 2020]{
\begin{tabular}{c|c}
    \hline
    \textbf{Country}   & \textbf{\thead{No.\ of\\ Distinct IPs}} \\
     \hline 
      Egypt   & 3195 \\
      Taiwan   & 1502 \\
      China  & 1373 \\
      Brazil  & 1074 \\
      France & 946 \\
      
      \hline
    \end{tabular}}
\qquad
\subfloat[][Sept.\ 6, 2020]{
\begin{tabular}{c|c}
    \hline
    \textbf{Country}   & \textbf{\thead{No.\ of\\ Distinct IPs}} \\
     \hline 
      Egypt   & 15293 \\
      Taiwan   & 1933 \\
      China  & 1238 \\
      Brazil  & 1019 \\
      France & 839 \\
      
      \hline
    \end{tabular}}

\end{table}

      

\subsection{The embedded devices incident}

\begin{table}[ht]
\centering
\caption{Top 5 ports (ranked by unique IPs) }
\label{table:top_ports_unique_nov}

\subfloat[][Nov.\ 26, 2020]{
\begin{tabular}{c|c}
    \hline
    \textbf{Port}   & \textbf{\thead{No.\ of\\ Distinct IPs}} \\
     \hline 
      23   & 61281 \\
      445   & 56971 \\
      59478  & 55741 \\
      80  & 48336 \\
      8080 & 46665 \\
      
      \hline
    \end{tabular}}
\qquad
\subfloat[][Nov.\ 27, 2020]{
\begin{tabular}{c|c}
    \hline
    \textbf{Port}   & \textbf{\thead{No.\ of\\ Distinct IPs}} \\
     \hline 
      50668   & 53753 \\
      445   & 53109 \\
      23  & 50763 \\
      80  & 44250 \\
      8080 & 43016 \\
      
      \hline
    \end{tabular}}

\end{table}

      

      

\begin{table}[ht]
\centering
\caption{Top 5 ports (ranked by total packets) }
\label{table:top_ports_tot_nov}

\subfloat[][Nov.\ 26, 2020]{
\begin{tabular}{c|c}
    \hline
    \textbf{Port}   & \textbf{\thead{Total \\ Packets}} \\
     \hline 
      80   & 1287486 \\
      8080   & 1268836 \\
      23  & 1113426 \\
      445  & 782743 \\
      59478 & 718157 \\
      
      \hline
    \end{tabular}}
\qquad
\subfloat[][Nov.\ 27, 2020]{
\begin{tabular}{c|c}
    \hline
    \textbf{Port}   & \textbf{\thead{Total \\ Packets}} \\
     \hline 
      80   & 1172944 \\
      8080   & 1156958 \\
      23  & 1009036 \\
      445  & 731203 \\
      50668 & 689067 \\
      \hline
    \end{tabular}}

\end{table}

      

      

\begin{figure}[t]
    \centering
    \includegraphics[width=3in]{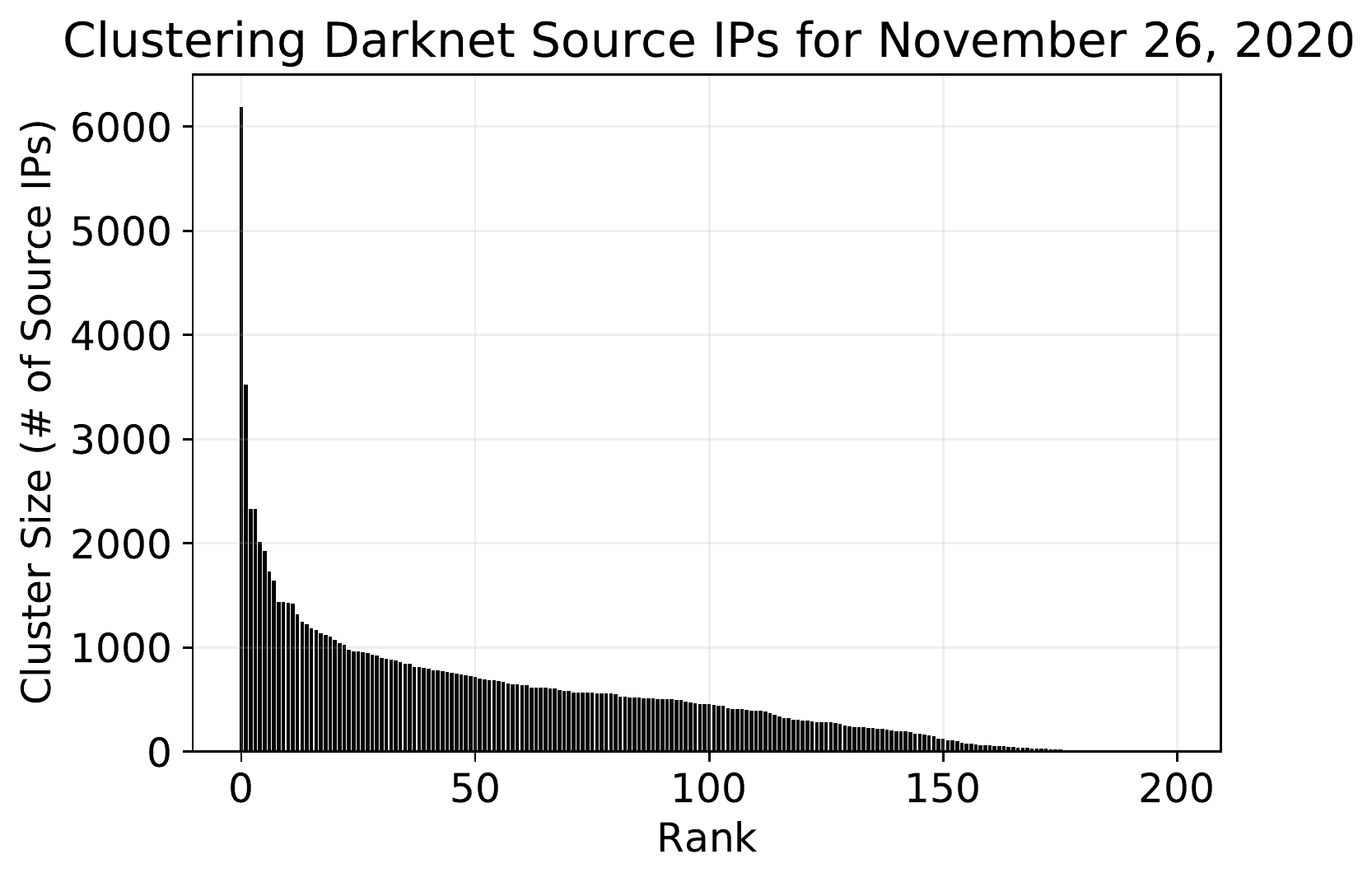}
    \caption{Distribution of cluster sizes for November 26.}
    \label{fig:scd26}
\end{figure}

\begin{figure}[t]
    \centering
    \includegraphics[width=3in]{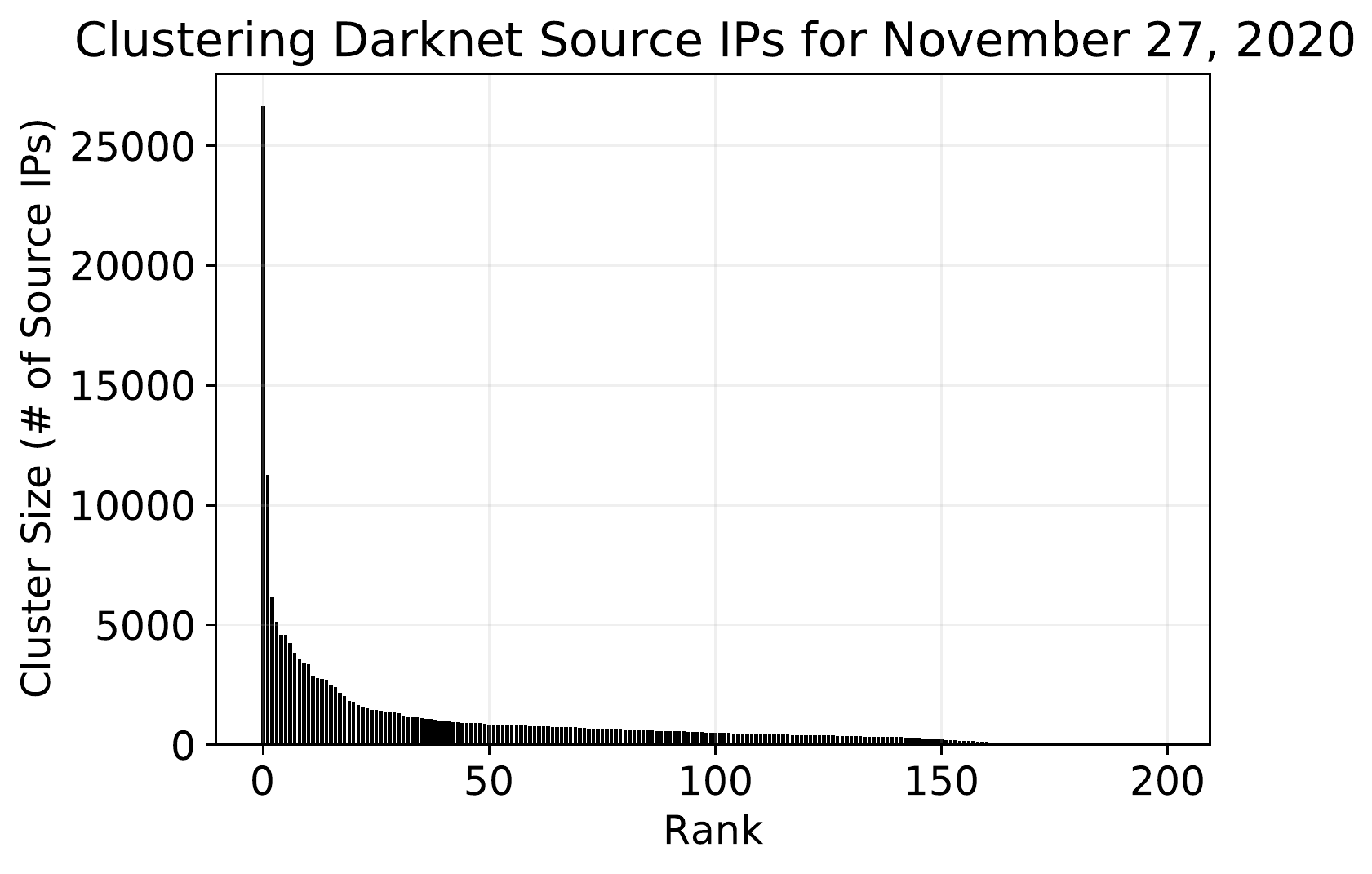}
    \caption{Distribution of cluster sizes for November 27.}
    \label{fig:scd27}
\end{figure}

The second case study we examine is attributed to the large EMD spike
detected on Nov.\ 27th (i.e., denoting a large difference in the clustering
outcomes between that day and the day before). Figures~\ref{fig:scd26} and~\ref{fig:scd27}
illustrate the distribution of the cluster sizes for these 2 days.
There are two extremely large clusters formulated on Nov.\ 27th 
with sizes 26662 (cluster-1) and 11262 (cluster-19). The largest cluster on Nov.\ 26th had only
about 6000 members (cluster-2). We here start to speculate that these 2 large clusters might be the culprits 
for the dramatic shift in the EMD distances, but we need to look
deeper into the clustering results. 

Figure~\ref{fig:emd} does not indicate that the change is associated with
any of the top-5 ports for that month. Similarly,
Table~\ref{table:top_ports_unique_nov} and \ref{table:top_ports_tot_nov} that include the top-5 ports zoomed into
both days do not provide any further insights. This illustrates that
``group by" operations (performed frequently by forensic analysts when trying
to troubleshoot an incident) might sometimes provide limited information. 
Further, when there are multiple categorical features (as is the case here, 
with categorical features being the ports scanned, protocols, Censys tags, Censys ports, countries, etc.)
and multiple numerical features, forensics analysis and result interpretation might be challenging.

\begin{figure}[t]
\centering

\begin{subfigure}[b]{.48\linewidth}\centering
\includegraphics[scale=0.5]{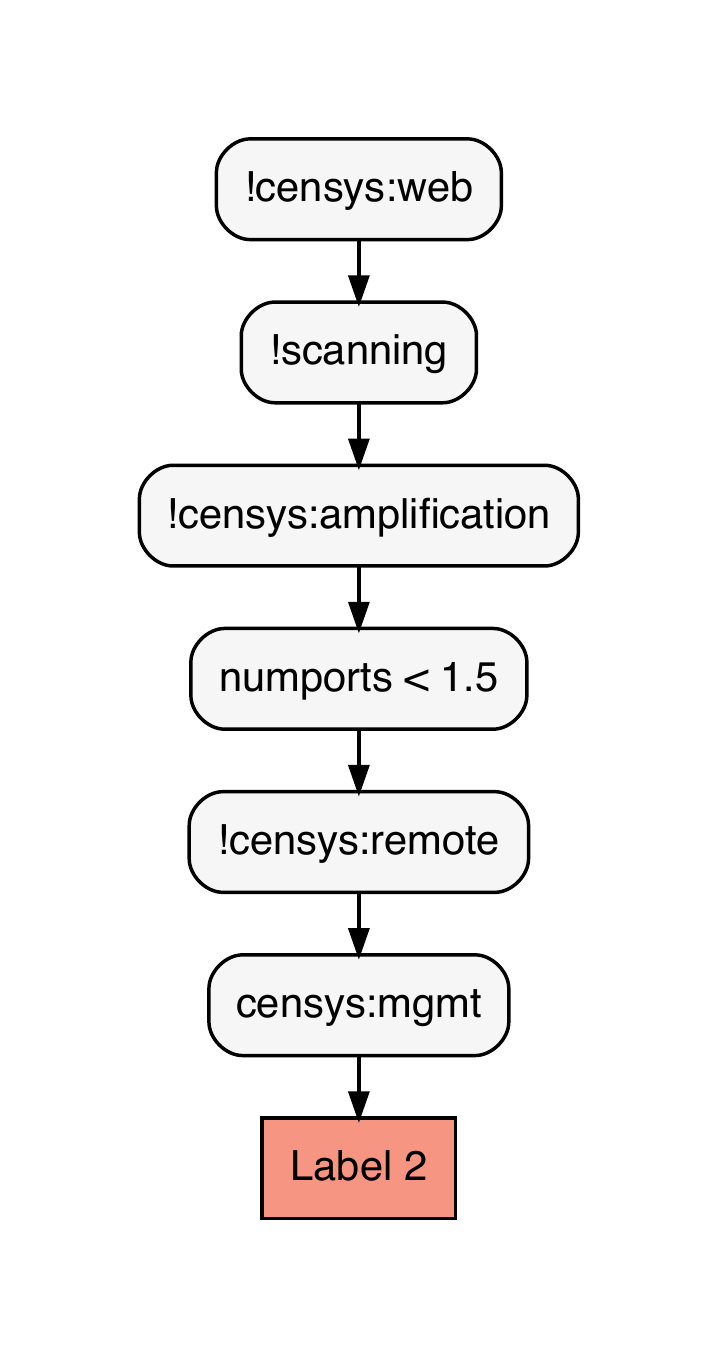}
\end{subfigure}
\begin{subfigure}[b]{.48\linewidth}\centering
\includegraphics[scale=0.5]{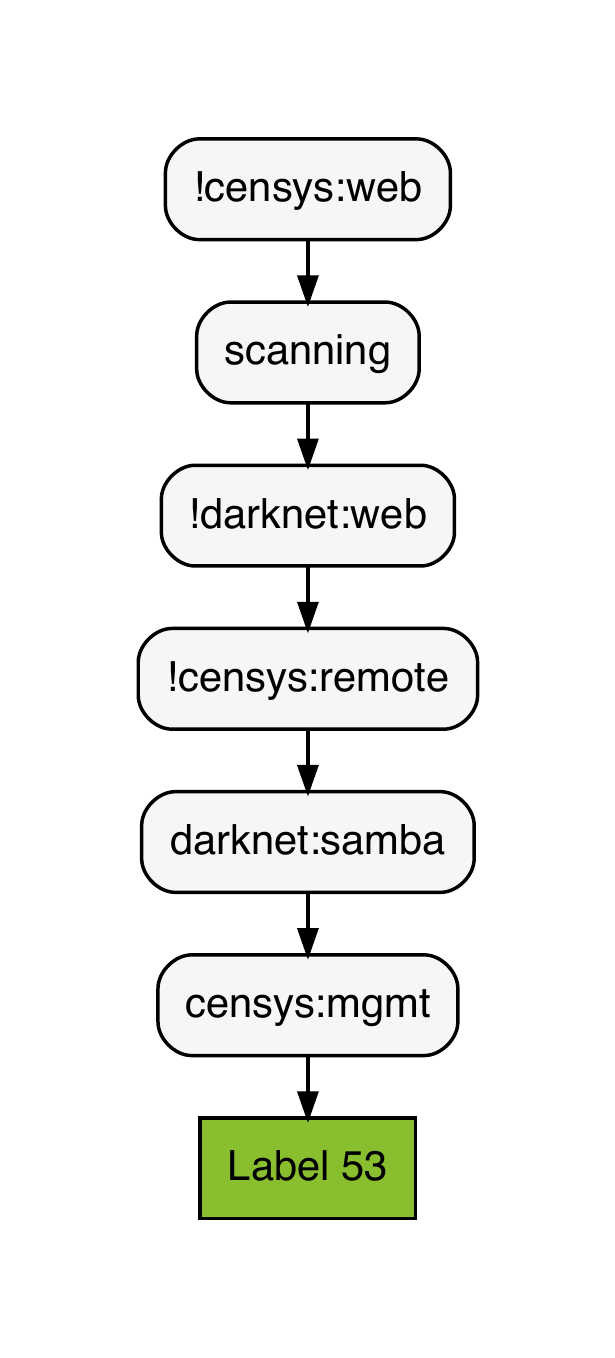}
\end{subfigure}

\vspace*{-12pt}

\begin{subfigure}[b]{.48\linewidth}\centering
\includegraphics[scale=0.5]{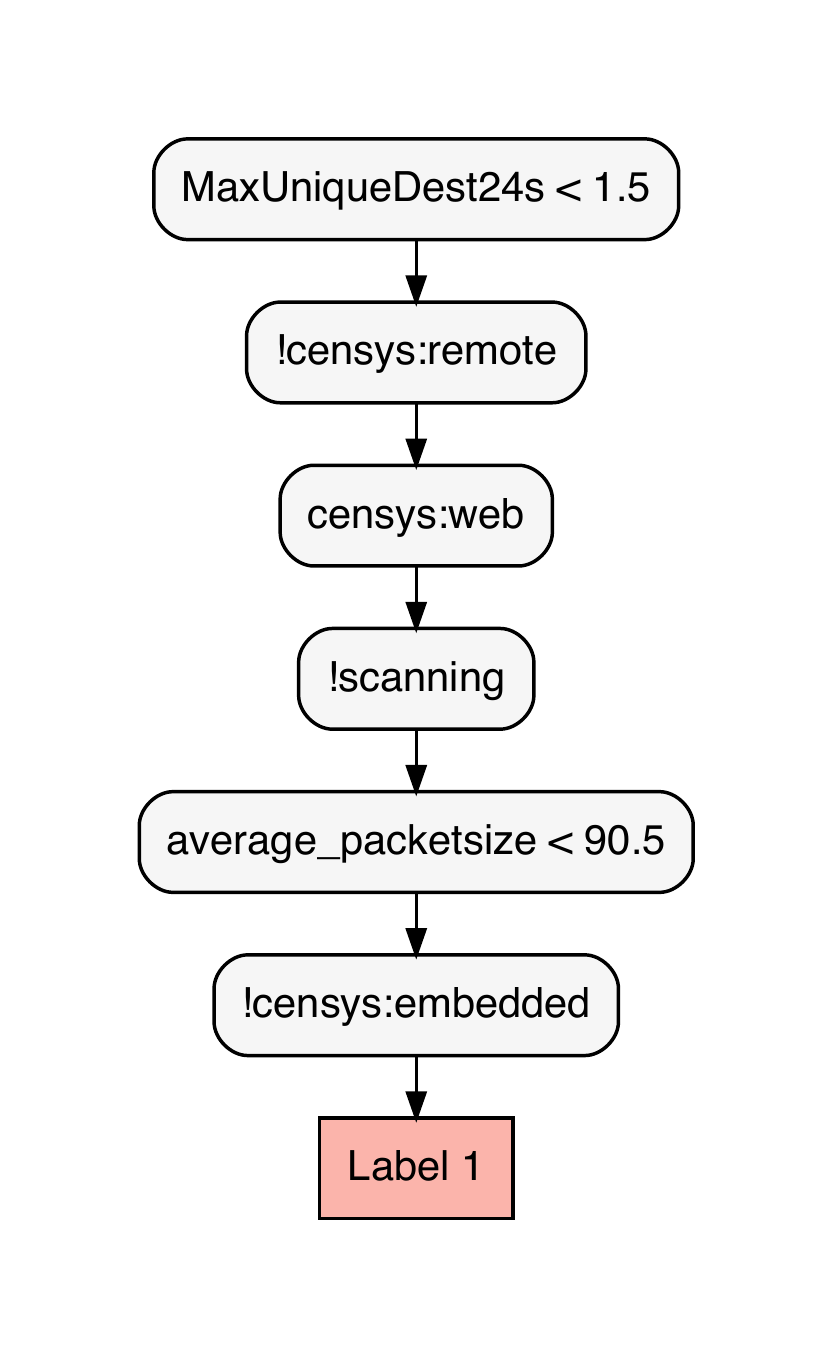}
\end{subfigure}
\begin{subfigure}[b]{.48\linewidth}\centering
\includegraphics[scale=0.5]{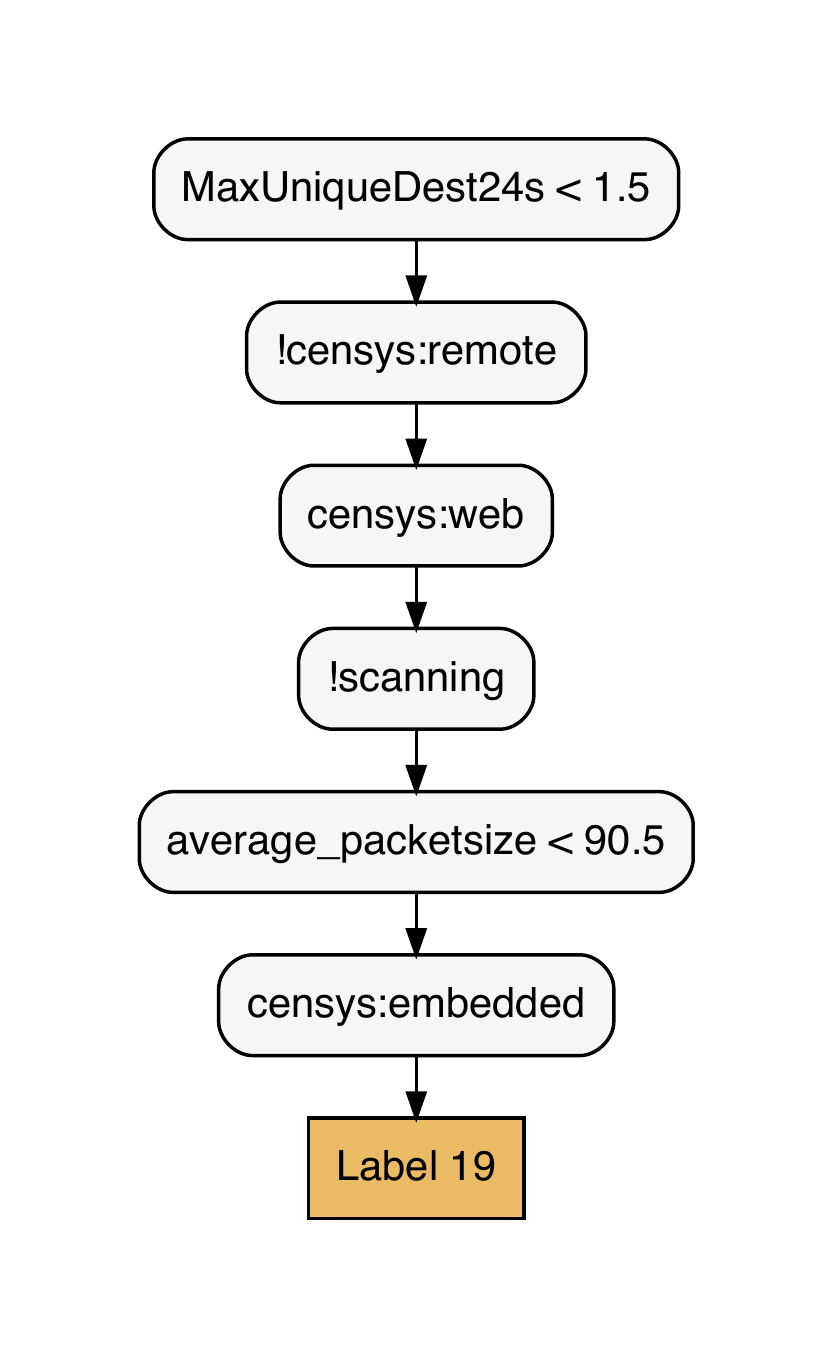}
\end{subfigure}

\vspace*{-12pt}

\caption{Decision Paths for the clusters of November 26th (top) and November 27th (bottom).}
\label{fig:dpaths}
\end{figure}

However, by reflecting on the decision tree paths depicted in Figure~\ref{fig:dpaths}
we start to gain some understanding about what might be responsible for the structural clustering
changes between the 2 days. The rule on the root for Nov. 26th immediately tells us
that the top 2 clusters are events that are not associated with “Censys:Web” (meaning that 
ports 80 or 443 were not found open when Censys scanned the scanners belonging in these clusters;
see Table~\ref{table:orion:tags}). 
At the same time, 
the decision rule “Censys:Web” is active for the top-2 clusters for Nov. 27th. Hence, the
top-2 very large clusters on Nov. 27th and the top-2 clusters on Nov. 26th
are characterized by a distinctive feature (i.e., the ``Censys:Web" grouping)
and this very likely explains the clustering differences observed.
Notably, the top cluster for Nov. 27th and the second-top cluster only differ on their last rule of 
their decision paths (i.e., the rule about “Censys: embedded”) which means
the clusters are very similar except when it comes with the ``embedded tag" assigned to them by Censys.

Examining Table~\ref{table:top_censys_ports}, which ranks the scanners
by their Censys-reported ``open ports", seems to validate that the culprits
for the Nov. 27th events are scanners with ports 80 and 443 open that became compromised
between the 2 days. Closer investigation of these scanners showed that the behavior was UDP-related
(shown also in the decision paths as ``!scanning", i.e., not TCP or ICMP scanning; see Table~\ref{table:orion:tags})
and the scanners were targeting a variety of randomly looking, high-value ephemeral ports. 
Further investigation is required to fully understand the intentions behind this event
that was originating from several countries and BGP routing prefixes. We note, though,
that only 606 source IPs from the top-cluster of Nov. 27th match with any of the
scanners from the previous day suggesting the event of Nov. 27th involved newly infected hosts.


\begin{table}[ht]
\centering
\caption{Top 5 Censys ports}
\label{table:top_censys_ports}

\subfloat[][Nov.\ 26, 2020]{
\begin{tabular}{c|c}
    \hline
    \textbf{Port}   & \textbf{\thead{No.\ of \\ Distinct IPs}} \\
     \hline 
      443   & 76585 \\
      80   & 65959 \\
      53  & 46299 \\
      22  & 39600 \\
      2000 & 38091 \\
      
      \hline
    \end{tabular}}
\qquad
\subfloat[][Nov.\ 27, 2020]{
\begin{tabular}{c|c}
    \hline
    \textbf{Port}   & \textbf{\thead{No.\ of \\ Distinct IPs}} \\
     \hline 
      80   & 138727 \\
      443   & 125603 \\
      22  & 57086 \\
      53  & 45862 \\
      7547 & 43274 \\
      \hline
    \end{tabular}}

\end{table}

      

      

\subsection{The SSH incident}

The last case study involves heavy SSH scanning associated primarily with devices/hosts
attributed to a US tier-1 ISP. Consulting Figure~\ref{fig:emd}, we see
a change-detection in the EMD chart between Jan. 6th and Jan. 7th. 
The SSH scanning activity peaks on Jan. 9th, as seen in the top-right panel
of Figure~\ref{fig:emd}, but the change in unique scanners between Jan. 6th and 7th
is not that obvious when visualizing that plot. Indeed, looking at the top-5
ports of Table~\ref{table:top_5_ports}, we see that port 22 (SSH) is 
absent (port 22 ranked top-44th on Jan. 6th and top-13th on Jan. 7th).

Nevertheless, the clustering approach is able to capture this onset of this
important event. Looking at Table~\ref{table:top_mirai_Jan6},
we observe that there was only a single cluster with port 22 as the
top port within the cluster. However, the clustering
structure changes significantly on Jan. 6th as evidenced by Table~\ref{table:top_mirai_Jan7}.
We see several SSH-related ports emerging and many of the clusters are associated with the CWMP
tag. This suggests that the infected devices might be ISP-managed modems that 
were susceptible and got compromised. 
We note here that, although not captured by the EMD detector,
a similar SSH incident occurred back on Nov. 6th, 2020 that involved the same tier-1 ISP.
In both cases we notified their ``network abuse" team about the incidents. 

\begin{figure}[ht]
    \centering
    \includegraphics[width=3.5in,trim=1cm 1cm 0 0,clip]{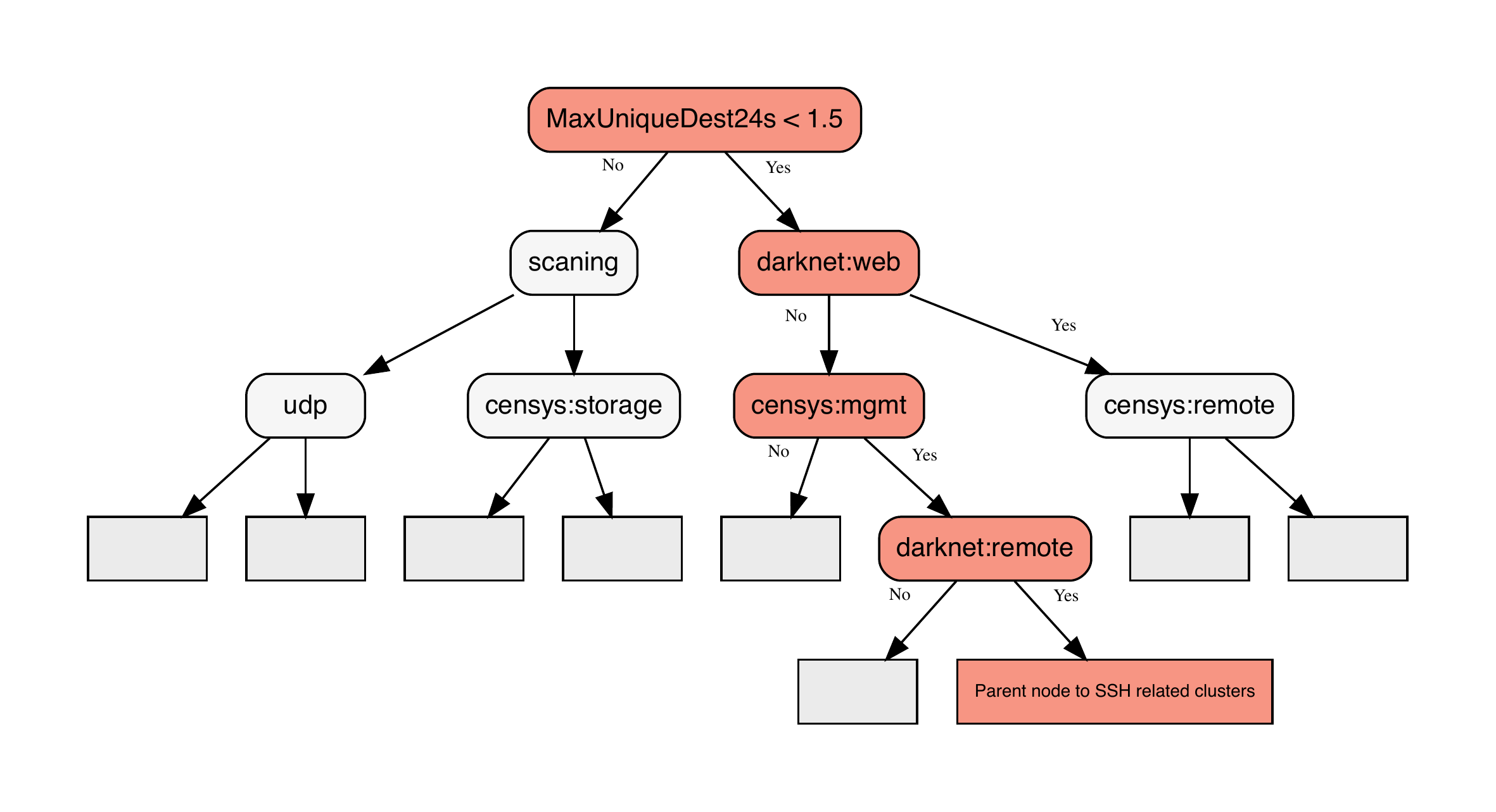}
    \caption{A common decision path that explains 5 out of 6 SSH related clusters on Jan.\ 7. The single path shows that the decision tree interpretation is able to capture the most important common characteristics of similar clusters. }
    \label{fig:cd26}
\end{figure}

Figure~\ref{fig:cd26} illustrates a pruned version of the optimal decision tree
obtained for Jan. 7th (the full tree is of depth 7). It again demonstrates the utility
of decision trees for clustering interpretation. 
Five out of six clusters associated with SSH activity are extensively explained by the decision tree path highlighted
with the red nodes that leads to Node 90 in the tree. 
That set of rules explain the following fraction of scanners (see
Table~\ref{table:top_mirai_Jan7}): for cluster-20: 2468/2647, for cluster-35: 5621/5739, 
for cluster-44 192/958, for cluster-66: 1367/1441, for cluster-74: 2/2023 and for cluster-105: 4233/4325.

\begin{table}[ht]
\centering
\caption{Top 5 ports}
\label{table:top_5_ports}

\subfloat[][Jan.\ 6, 2020]{
\begin{tabular}{c|c}
    \hline
    \textbf{Port}   & \textbf{\thead{No.\ of \\ Distinct IPs}} \\
     \hline 
      445   & 54946 \\
      51974   & 52268 \\
      23  & 44244 \\
      80  & 40599 \\
      8080 & 38950 \\
      
      \hline
    \end{tabular}}
\qquad
\subfloat[][Jan.\ 7, 2020]{
\begin{tabular}{c|c}
    \hline
    \textbf{Port}   & \textbf{\thead{No.\ of \\ Distinct IPs}} \\
     \hline 
      445   & 54791 \\
      63570   & 54067 \\
      23  & 45265 \\
      80  & 43097 \\
      8080 & 41201 \\
      \hline
    \end{tabular}}

\end{table}

      

      

\begin{table}[htb]
    \centering
    \caption{Only one small SSH-related cluster on Jan.~6, 2021}
    \footnotesize
    \label{table:top_mirai_Jan6}
    \resizebox{\linewidth}{!}{\begin{tabular}{| c |c || c | c || c | c || c |}
    \hline
    \textbf{\thead{Cluster\\ Label}}   & \textbf{\thead{Cluster\\ Size}} & \textbf{\thead{Top\\ Ports}}   & \textbf{\thead{Top Port\\ Fraction}} & \textbf{\thead{Top\\ Tags}} & \textbf{\thead{Top Tag\\ Fraction}} &
    \textbf{\thead{Mirai\\ Coverage}}\\
     \hline 
      96 & 703        &   22    & 0.34 & http & 0.28 & 0.36\\
      \hline
      
    \end{tabular}}
\end{table}

\begin{table}[htb]
    \centering
    \caption{Several SSH-related clusters emerge on Jan.~7, 2021}
    \footnotesize
    \label{table:top_mirai_Jan7}
    \resizebox{\linewidth}{!}{\begin{tabular}{| c |c || c | c || c | c || c |}
    \hline
    \textbf{\thead{Cluster\\ Label}}   & \textbf{\thead{Cluster\\ Size}} & \textbf{\thead{Top\\ Ports}}   & \textbf{\thead{Top Port\\ Fraction}} & \textbf{\thead{Top\\ Tags}} & \textbf{\thead{Top Tag\\ Fraction}} &
    \textbf{\thead{Mirai\\ Coverage}}\\
     \hline 
      35 & 5739        &   22    & 0.99 & cwmp & 0.95 & 0.00\\
      105 & 4325        &   22    & 0.98 & cwmp & 0.84 & 0.01\\
      20 & 2647        &   22    & 0.95 & embedded & 0.33 & 0.01\\
      74 & 2023        &   22    & 0.93 & https & 0.49 & 0.01\\
      66 & 1441        &   22    & 0.92 & cwmp & 0.75 & 0.08\\
      44 & 958        &   22    & 0.45 & http & 0.22 & 0.26\\
      \hline
      
    \end{tabular}}
\end{table}

\section{Discussion and Future Directions}
\label{future}

Security analysts have several opportunities to
leverage this near-real-time clustering approach proposed here. 
First, the high
intensity scanners identified can be 
combined with other scanning data collected by an organization
to improve their network's risk assessment.
This is important because data collected by an
enterprise may not be able to easily identify 
high-intensity scanners, as we discussed in section
\ref{sec:DarknetObservabilities}.  Furthermore,
the ports being scanned by high-intensity scanners
can reveal information about potential vulnerabilities
that may be exploited by adversaries.  Subsequently,
enterprise network security teams can be more proactive
in revising risk assessment based on this new information
and mitigation strategies
for their network.
Second, Indicators of Compromise (IoC) generated by an
organization's Intrusion Detection System can
be filtered and prioritized using the clusters 
generated.  

One of the important directions of this research
is to extend the discussed clustering framework into
an automated multi-pass clustering.  Due to the
high speed of scanning data arriving at the Darknet
and the 
high dimensionality of its feature space, the size
of clusters generated can still be large, which 
often includes scanners 
with different characteristics. For example, a
high-intensity scanner that scans only a small number
of ports may be grouped with other scanners 
that scan similar ports, but with much lower
intensity.  Hence, an automated multi-pass clustering
of the proposed framework is needed to enable 
further clustering of clusters based on its size
and/or the entropy of critical features (e.g., ports scanned). 


\section{Conclusion}
\label{sec:conclusion}

This paper introduces a novel unsupervised 
clustering-based approach to characterizing Darknet scanning activity
and its evolution over time.
We exploit important information about the behaviors of Darknet scanners 
(e.g., the set of ports scanned by the scanner as well as information provided
by \texttt{Censys.io} about the set
of services open at the scanning IP) to encode the data. 
Such information is critical for characterizing 
Darknet events and can reveal vulnerabilities targeted by nefarious users. 
The resulting heterogeneity of the data features, the high dimensionality 
of the data, the need to cope with potentially non-linear 
interactions between features present challenges that are largely 
unaddressed by 
existing clustering-based approaches to analyses of Darknet data. 
Hence, we proposed a deep representation learning approach to 
clustering high dimensional data from network scanners. We
also employed optimal classification trees for result interpretation
and utilized the clustering outcomes as ``signatures`` to help
us detect structural Darknet changes over time.


\section*{Acknowledgements}

This research is partially 
supported by a U.S. Department of Homeland Security Grant through 
Center for Accelerating Operational Efficiency (CAOE).
The views and conclusions contained herein are those of the authors and should not be interpreted as  representing the official policies, either expressed or implied, of the U.S. Department of Homeland Security.



\bibliographystyle{plain}
\bibliography{refs,darknet,malware,deepclustering}







\end{document}